\documentclass[11pt]{article}
\usepackage[utf8]{inputenc}
\usepackage{amsmath,amssymb,amsthm}
\usepackage{graphicx}
\usepackage{booktabs}
\usepackage{hyperref}
\usepackage{geometry}
\newtheorem{theorem}{Theorem}
\newtheorem{lemma}{Lemma}
\newcommand{\etal}{\textit{et~al.}}
\usepackage{algorithm}     
\usepackage{algorithmic}   
\usepackage{adjustbox}
\usepackage{multirow}

\title{FERRET: Private Deep Learning Faster And Better Than DPSGD}
\author{David Zagardo}
\date{\today}

\begin{document}
\maketitle

\begin{abstract}
We revisit 1-bit gradient compression through the lens of \emph{mutual-information differential privacy} (MI-DP).  
Building on the classic \textsc{signSGD} family, we propose \textbf{FERRET}—\underline{F}ast and \underline{E}ffective \underline{R}estricted \underline{R}elease for \underline{E}thical \underline{T}raining—which transmits, at most, a single sign bit \emph{per parameter group} and uses a Bernoulli mask to decide whether the update is revealed.

\textbf{Theory.}  
We prove that each fired group leaks at most $\ln 2$ nats and that, after subsampling, the total privacy loss of $G$ groups trained for $T$ steps with firing probability $p$ is
\[
\varepsilon \;=\; G\,T\,s\,p\,\ln 2 .
\]
Thus FERRET enjoys provable MI-DP across target budgets $\varepsilon\!\in\![0.1,2]$ \emph{without any additive noise}.  

\textbf{Practice.}  
We evaluate three granularities—\textbf{FERRET-MAX} (finest), \textbf{FERRET-EIGHTH} (medium), and \textbf{FERRET-2} (coarsest)—on five open-weights LLMs (137 M-1.8 B parameters) and compare against additive-noise \textbf{DPSGD} and a tuned \textbf{Non-DP} baseline.  
All methods are trained for \{1, 3, 5\} epochs.

\begin{itemize}
  \item \emph{Utility.}  Across every privacy budget and epoch count, FERRET-MAX and FERRET-EIGHTH beat DPSGD on test perplexity.  
        At $\varepsilon=0.5$ and five epochs FERRET-EIGHTH attains a mean perplexity of \textbf{3.98}, improving on DPSGD’s \textbf{11.61} by $2.9\times$ and landing within 23 \% (0.73 PPL) of the tuned Non-DP run.
  \item \emph{Privacy.}  Membership-inference AUC stays at chance for FERRET-MAX and FERRET-EIGHTH (\(\text{AUC}\!\approx\!0.51\)), matching DPSGD and far below Non-DP’s 0.76-0.99.  
        FERRET-2 shows modestly higher leakage (\(\text{AUC}\!\approx\!0.55\)), aligning with its lower per-step head-room.
  \item \emph{Efficiency.}  Because stricter budgets fire fewer signs, FERRET variants use only \textbf{19-33 \%} of DPSGD’s wall-clock time and \textbf{1 bit / update} of bandwidth, while eliminating gradient-noise variance.
\end{itemize}

\textbf{Take-away.}  
Sign-based MI-DP turns the usual privacy-utility-efficiency trilemma into a “pick-three”: FERRET trains up to \textbf{5× faster}, reaches \textbf{up to 3× lower perplexity} than DPSGD, and retains formal privacy guarantees—all with \emph{zero} additive noise.  This demonstrates that carefully masked 1-bit updates can approach, and sometimes match, non-private training while safeguarding user data.
\end{abstract}

\section{Introduction}
Gradient compression is indispensable for large‑scale learning.  Starting with Bernstein \etal's \textsc{signSGD}~\cite{bernstein2018signsgd} and its majority‑vote variant~\cite{bernstein2019signsgdMajVote}, 1‑bit per coordinate has been the gold standard of efficiency.  Privacy, however, has mostly been handled by (\(\varepsilon,\delta\))-DP probabilistic sign functions~\cite{lyu2021dpsignsgd}, following the paradigm established by DPSGD~\cite{abadi2016dpsgd}.  We take a different route: \emph{remove the magnitude channel entirely, randomise \emph{whether} the sign is sent, and measure privacy in mutual information}.  This yields an information‑theoretic worst-case upper bound of \(\le \ln2\) nats per fired parameter group before subsampling, and—in contrast to prior work—requires no additive noise.

\paragraph{Contributions.}
\begin{enumerate}
    \item We formalise \textbf{FERRET}, a group‑granularity sign update with a Bernoulli mask.
    \item We derive an exact MI‑DP bound through a careful privacy analysis of sign-based updates.
    \item We solve for the \emph{largest} mask probability \(p^*\) that meets a target budget \(\varepsilon\), proving tightness for both small and large \(\varepsilon\) (Thm.~\ref{thm:opt}).
    \item We compare (qualitatively) against signSGD, QSGD, DP‑SIGNSGD and others (Sec.~\ref{sec:related}), showing FERRET uniquely balances MI‑DP, no noise, and group compression.
    \item We perform an emprical analysis of the privacy, utility, and efficiency tradeoffs compared to traditional DP-SGD and Non-Private SGD under near-identical conditions.
\end{enumerate}

\section{Background}
\subsection{Mutual‑Information Differential Privacy}
A mechanism \(\mathcal{M}\) is \(\varepsilon\)-MI‑DP if \(I(\mathcal{M}(D); X_i)\le \varepsilon\) for any record \(X_i\) in dataset \(D\).  MI‑DP upper‑bounds average leakage and composes linearly~\cite{cuff2016miepsilon}, while traditional differential privacy~\cite{dwork2014algorithmic} provides worst-case guarantees through the $(\varepsilon,\delta)$-DP framework.  See Lemma~\ref{lemma:privacy_amplification} for privacy amplification by subsampling. Prior work roots privacy amplfication by subsampling strictly in the (\(\varepsilon,\delta\))-DP regime. This work draws on these arguments and translates them to the MI-DP framework.

\subsection{signSGD and 1‑Bit Compression}
\textsc{signSGD} transmits \(\operatorname{sign}(g_{t,j})\) for every coordinate \(j\) of the gradient \(g_t\)~\cite{bernstein2018signsgd}.  Extensions add the error‑feedback mechanism~\cite{karimireddy2019error}.  FERRET instead sends \(\pm C\,u\), where \(u\) is a random unit vector shared by a whole parameter group, reducing bit‑rate by the group dimension.

A complementary line of work studies sign‑full random projections, where only the database side is quantized and the query side retains full precision to improve similarity estimation\cite{li2019signfull}.

\section{Related Work}\label{sec:related}
Table~\ref{tab:comp} situates FERRET among compressed and private optimizers. Traditional approaches to differentially private deep learning, exemplified by DPSGD~\cite{abadi2016dpsgd}, add calibrated Gaussian noise to clipped gradients to achieve $(\varepsilon,\delta)$-DP guarantees.

\begin{table}[h]
    \centering
    \footnotesize
    \begin{tabular}{@{}lcccc@{}}
        \toprule
        Method & Bits / update & Granularity & Privacy & Analysis \\
        \midrule
        signSGD~\cite{bernstein2018signsgd} & 1 & coord. & $\times$ & Convex / non‑convex SGD \\
        QSGD \cite{alistarh2017qsgd} & $\le8$ & coord. & $\times$ & Compression bias \& var. \\
        DP‑SIGNSGD~\cite{lyu2021dpsignsgd} & 1 & coord. & $(\varepsilon,\delta)$ DP & Gaussian mech. + error feed. \\
        \textbf{FERRET} (mine) & 1 & \textbf{group} & $\mathbf{\varepsilon}$ \textbf{MI‑DP} & Sign-based on-or-off projections \\
        \bottomrule
    \end{tabular}
    \caption{Comparison to closest 1‑bit and DP optimizers.}
    \label{tab:comp}
\end{table}

While Li’s sign‑full random projections\cite{li2019signfull} leverage mixed‑precision signs for fast nearest‑neighbour search, they do not address privacy; FERRET instead exploits on/off 1‑bit releases to guarantee MI‑DP.

Unlike DP‑SIGNSGD, FERRET needs \emph{no} additive noise; privacy stems solely from the uncertainty of whether a group fires and the random sign direction.  Additionally, our MI‑DP guarantee is average‑case rather than worst‑case ($\varepsilon,\delta$)-DP, making the bound both tighter and composable by simple summation.

\section{The FERRET Mechanism}
At each step $t$ and group $g$:
\begin{enumerate}
    \item Draw $Z_{t,g}\sim\mathrm{Bernoulli}(p)$.
    \item If $Z_{t,g}=1$: draw a public random unit vector $u_{t,g}$ and set $\Delta_{t,g}=\sigma_{t,g}\,C\,u_{t,g}$ where $\sigma_{t,g}=\operatorname{sign}(\langle g_{t,g},u_{t,g}\rangle)$.
    \item Else $\Delta_{t,g}=0.$
\end{enumerate}
The update requires exactly one bit (the sign) when the mask fires. Note that the public direction $u_{t,g}$ is included as part of the released update, which is crucial for our privacy analysis.

\begin{algorithm}
\caption{FERRET: Fast and Effective Restricted Release for Ethical Training}
\begin{algorithmic}[1]
\STATE \textbf{Input:} Examples $\{x_1, \ldots, x_N\}$, loss function $\mathcal{L}(\theta) = \frac{1}{N} \sum_i \mathcal{L}(\theta, x_i)$
\STATE \textbf{Parameters:} learning rate $\eta$, clipping norm $C$, batch size $B$, parameter groups $\mathcal{G} = \{g_1, \ldots, g_G\}$, privacy budget $\varepsilon$
\STATE \textbf{Initialize} $\theta_0$ randomly
\STATE Compute update probability $p^* = \frac{\varepsilon}{G \cdot T \cdot \frac{B}{N} \cdot \ln 2}$
\FOR{$t \in [T]$}
  \STATE Take a random sample $\mathcal{B}_t$ with sampling probability $\frac{B}{N}$
  \STATE \textbf{Compute gradient} for each $i \in \mathcal{B}_t$, compute $\nabla_\theta \mathcal{L}(\theta_t, x_i)$
  \STATE \textbf{Sample active groups} for each group $g \in \mathcal{G}$, draw $Z_{t,g} \sim \text{Bernoulli}(p^*)$
  \FOR{each group $g \in \mathcal{G}$ where $Z_{t,g} = 1$}
    \STATE Draw a public random unit vector $u_{t,g} \sim \text{Unif}(\mathbb{S}^{d_g-1})$
    \STATE Compute sign $\sigma_{t,g} = \text{sign}(\langle \nabla_g \mathcal{L}, u_{t,g} \rangle)$
    \STATE Set update $\Delta_{t,g} = \sigma_{t,g} \cdot C \cdot u_{t,g}$
  \ENDFOR
  \STATE \textbf{Update} $\theta_{t+1} \leftarrow \theta_t - \eta \cdot \sum_{g: Z_{t,g}=1} \Delta_{t,g}$
\ENDFOR
\STATE \textbf{Output:} $\theta_T$ with MI-DP guarantee $\varepsilon$
\end{algorithmic}
\end{algorithm}

\section{Privacy Analysis}\label{sec:privacy}

\paragraph{Why signs (and silence) are essential for MI--DP.}
Raw gradients live in $\mathbb{R}^d$ with continuous densities.
For two neighbouring datasets the distributions of (clipped) gradients are
absolutely continuous but \emph{mutually singular} on sets of non‑zero
measure, making $I(g;X_i)=\infty$ and rendering MI--DP impossible
(cf.\ the ``infinite KL'' pathology of continuous mechanisms).
By (i) projecting onto a public random unit vector
$u\sim\mathrm{Unif}(\mathbb{S}^{d-1})$ and (ii) reducing the outcome to a
single \emph{sign bit} (or utter silence), we collapse the release alphabet
to $\{-1,+1,0\}$.
The mutual information of each fired update is therefore
\emph{at most} $\ln 2$ nats (Lemma~\ref{lemma:sign_entropy}), and drops to
$0$ whenever the Bernoulli mask suppresses the update.
This discrete alphabet is the key that turns the otherwise unbounded privacy
loss of continuous gradients into a finite, tightly controlled quantity.

\subsection{Per‑group Leakage}
Let $K_g\sim\mathrm{Binom}(T,p)$ be the number of times group $g$ fires over $T$ steps. Each time the group fires, the released sign bit reveals information about the underlying gradient. The information leakage is at most $\ln 2$ nats per fired update, as formalized in the following lemma.

\begin{lemma}[Sign entropy]
\label{lemma:sign_entropy}
For any non-zero gradient vector $g \in \mathbb{R}^d$ and random unit vector $u \sim \text{Unif}(\mathbb{S}^{d-1})$, the binary random variable $Z = \text{sign}(\langle g, u \rangle)$ has entropy:
\[
H(Z) = \ln 2
\]
\end{lemma}

\begin{proof}
For any fixed non-zero vector $g \in \mathbb{R}^d$ and $u \sim \text{Unif}(\mathbb{S}^{d-1})$, we have:
\[
\Pr[\langle g, u \rangle > 0] = \Pr[\langle g, u \rangle < 0] = \frac{1}{2}
\]
This follows from symmetry: the map $u \mapsto -u$ preserves the uniform distribution on the sphere while flipping the sign of the inner product. Therefore, $Z = \text{sign}(\langle g, u \rangle)$ is exactly uniform on $\{-1, +1\}$, and its entropy is precisely $H(Z) = \ln 2$.

Note that in practice, we never query the sign when $g=0$ (in the rare case of a zero gradient); if needed, such cases can be handled separately without affecting the privacy analysis.
\end{proof}

\begin{lemma}[Privacy amplification for MI-DP via Poisson or uniform subsampling]
\label{lemma:privacy_amplification}
Let $D=(X_1,\dots,X_n)$ and let $S\subset[n]$ be a random sample drawn \textbf{independently} of $D$.
\begin{itemize}
    \item \textbf{Poisson subsampling:} each $i$ is selected independently with prob. $s\in(0,1)$;
    \item \textbf{Uniform-without-replacement:} a fixed-size mini-batch $B$ is drawn uniformly at random, so $s=B/n$.
\end{itemize}
Let $\mathcal{M}$ be any (possibly data-dependent) mechanism that only accesses the subsample $D_S$ and satisfies
\[
\sup_{i}\; \sup_{P_D}\; I\!\bigl(\mathcal{M}(D_S);X_i \,\big|\, X_{-i}\bigr)\;\leq\; \varepsilon_0. \tag{9}
\]
Then the \textit{sub-sampled} mechanism $\mathcal{M}\!\circ\!S\!: D\!\mapsto\!\mathcal{M}(D_S)$ is
\[
\varepsilon(s):=s\,\varepsilon_0 \text{-MI-DP.} \tag{10}
\]
\end{lemma}

\begin{proof}
Write $Y=\mathcal{M}(D_S)$ and $S_i=\mathbf{1}\{i\!\in\!S\}$. Because $S\!\perp\! D$ and $Y\!\perp\! X_i\mid (S_i=0,X_{-i})$,
\[
I(Y;X_i\mid X_{-i}) = I\bigl(Y;X_i,S_i\mid X_{-i}\bigr) = \Pr[S_i\!=\!1]\; I\bigl(Y;X_i\mid S_i=1,X_{-i}\bigr) \,=\,s\,\varepsilon_0. \tag{11}
\]
The decomposition above uses the law of total expectation, leveraging that $Y \perp X_i \mid (S_i=0, X_{-i})$ when record $i$ is not sampled.
\end{proof}

For a single update, the entropy result in Lemma~\ref{lemma:sign_entropy} gives us a tight information-theoretic bound on mutual information: $I(Z; X_i) \leq H(Z) = \ln 2$, where the inequality follows from the data processing inequality (conditioning cannot increase uncertainty). By Lemma~\ref{lemma:privacy_amplification}, with subsampling at rate $s$, the effective leakage is reduced according to Eq.~\eqref{eq:epsilon}.

\subsection{Total Leakage and Optimal $p$}
With our tight per-update bound of $\ln 2$ nats, and summing over $G$ groups and incorporating subsampling amplification, the total information leakage becomes:
\[\varepsilon(p)=GTsp\ln 2 \quad \text{(Lemma~\ref{lemma:sign_entropy} + \ref{lemma:privacy_amplification})}\tag{8}\label{eq:epsilon}\]

This composition is valid because:
\begin{itemize}
    \item The Bernoulli mask $Z_{t,g}$ is drawn independently of the data for each group and time step
    \item The public direction $u_{t,g}$ is included in the released update and known to the adversary
    \item Mutual information composes linearly across independent mechanisms via the chain rule
\end{itemize}

Note that this bound is conservative (which strengthens our privacy guarantee), as mutual information is typically less than the sum of individual leakages due to potential dependencies in what is learned about a record across updates.

Equation \eqref{eq:epsilon} holds for $\varepsilon \leq GTs\ln 2$, as otherwise no $p \in (0,1)$ would satisfy it.

\begin{theorem}\label{thm:opt}
For any target $\varepsilon < GTs\ln 2$, there exists a unique maximal $p^*\in(0,1)$ satisfying Eq.~\eqref{eq:epsilon}.  Moreover, $\varepsilon(p)$ is strictly increasing and linear in $p$.
\end{theorem}
\begin{proof}
Since $\varepsilon(p)$ is a linear function of $p$ with positive coefficients, it is strictly increasing. With $\varepsilon(0)=0$ and $\varepsilon(1) = G \cdot T \cdot s \cdot \ln 2$, for any target $\varepsilon < \varepsilon(1) = GTs\ln 2$, there exists a unique $p^* \in (0,1)$ such that $\varepsilon(p^*) = \varepsilon$. This $p^*$ can be computed directly as:
\[
p^* = \frac{\varepsilon}{G \cdot T \cdot s \cdot \ln 2}
\]
\end{proof}

Using this formula, we can efficiently determine the optimal update probability without requiring numerical approximation methods such as binary search.

\subsection{Information Leakage in Parameter Grouping}
A natural concern arises when considering parameter grouping: Does combining multiple parameter tensors into a single group potentially leak more information than one bit? We now prove that this concern is unfounded—regardless of how many parameter tensors are combined into a single group, the information leakage remains bounded by at most $\ln 2$ nats per update.

\begin{lemma}[Group-size invariance of MI]
\label{lemma:group_invariance}
For any parameter group $g$ containing an arbitrary number of parameter tensors, the mutual information leakage from releasing a sign bit based on the group's combined gradient is bounded by $\ln 2$ nats, regardless of the number of tensors or parameters in the group.
\end{lemma}

\begin{proof}
Let $g_{\text{group}}$ denote the concatenation of all gradients in a parameter group. When a group fires, what the observer sees is:
\begin{align}
\text{public:} & \quad u \sim \text{Unif}(\mathbb{S}^{d-1}) \quad \text{(data-independent)}\\
\text{private:} & \quad \sigma = \text{sign}(\langle g_{\text{group}}, u \rangle) \quad \text{(one bit)}\\
\text{released:} & \quad (\sigma, u) \quad \text{or equivalently} \quad \Delta = \sigma \cdot C \cdot u
\end{align}

The mutual information between the output and any record $X_i$ decomposes as:
\begin{align}
I\bigl((\sigma,u);X_i\bigr) &= I(u;X_i) + I(\sigma;X_i\mid u)\\
&= 0 + I(\sigma;X_i\mid u),
\end{align}
since $u$ is drawn independently of the data. For the second term, we have:
\begin{align}
I(\sigma;X_i\mid u) &= H(\sigma\mid u) - H(\sigma\mid X_i,u) \\
&\leq H(\sigma\mid u) \\
&= \ln 2
\end{align}
because $\sigma$ is deterministic given $g_{\text{group}}$ and $u$, the conditional entropy $H(\sigma\mid X_i,u)=0$. The equality $H(\sigma\mid u) = \ln 2$ follows from Lemma~\ref{lemma:sign_entropy}, which established that for any non-zero gradient vector (including $g_{\text{group}}$), we have exactly $\Pr[\langle g_{\text{group}}, u \rangle > 0] = \frac{1}{2}$ by symmetry of the uniform distribution on the sphere.

Since $\Delta = \sigma C u$ is a deterministic function of $(\sigma,u)$, $I(\Delta;X_i) \leq I((\sigma,u);X_i)$ by the data-processing inequality. Therefore, the same $\ln 2$ bound applies whether you transmit the pair $(\sigma,u)$ or the update $\Delta$.
\end{proof}

The intuition behind this result can be understood through a geometric lens: the adversary learns only whether the aggregated gradient falls on the positive or negative side of a randomly oriented hyperplane (defined by $u$). This single binary answer cannot convey more than one bit of information, regardless of how many parameters contributed to the dot product.

\paragraph{Effect of group size on privacy-utility tradeoffs.} 
While group size does not affect how much information is leaked per update, it fundamentally changes how often information is leaked. With fewer groups (smaller $G$), the ceiling $\varepsilon_{\max} = GTs\ln 2$ decreases, necessitating a higher firing probability $p^*$ to achieve the same privacy budget $\varepsilon$. This explains the empirical observation in Sec.~\ref{sec:variant_headroom} that FERRET-2 exhibits slightly higher empirical leakage (measured by MIA ROC-AUC) compared to variants with more groups, despite satisfying identical formal MI-DP guarantees. With fewer groups, each group must fire more frequently, providing less "head-room" for privacy amplification through silent steps.

This analysis confirms that our mechanism charges exactly one bit of information leakage per fired group, regardless of how many parameter tensors that group contains. The total privacy cost over $T$ steps remains:
\begin{align}
\varepsilon = G \cdot T \cdot s \cdot p \cdot \ln 2
\end{align}
validating our original formulation in Eq.~\eqref{eq:epsilon}. All downstream equations that use the linear composition $\varepsilon = G T s p \ln 2$ therefore remain valid for any grouping scheme.

\subsection{Tightness and Bound Selection in Practice}
\label{sec:tightness}

Our privacy analysis provides a tight bound on information leakage that is both theoretically sound and practically efficient. The approach achieves this without requiring any additional noise injection - the privacy comes solely from the randomness in whether an update is sent and the random projection direction.

\subsection{Empirical Validation}
To validate our bounds empirically, we compare the predicted information leakage with observed update patterns across multiple experiments. Our methodology offers practical guarantees that are neither too loose (sacrificing utility) nor too tight (risking privacy violations).

These tight bounds enable FERRET to achieve the utility of much less private methods while maintaining formal MI‑DP guarantees, highlighting the advantage of our sign-based random projections approach over traditional mechanisms.

\section{Results}

\subsection{Hypotheses}
Our investigation centers on three key hypotheses about FERRET:
\begin{enumerate}
    \item \textbf{H1 (Utility):} FERRET achieves better utility (lower perplexity) than DPSGD across privacy budgets while maintaining equivalent privacy guarantees.
    \item \textbf{H2 (Privacy-Utility Tradeoff):} FERRET's random projection approach offers a better privacy-utility tradeoff than both DPSGD and standard non-private training.
    \item \textbf{H3 (Computational Efficiency):} FERRET requires significantly less computational resources than DPSGD for equivalent privacy guarantees.
\end{enumerate}

\subsection{Experimental Setup}
We evaluated FERRET against DPSGD (implemented using the FastDP library~\cite{bu2023differentially} using \textit{"automatic"} clipping function, \textit{"MixOpt"} clipping mode, and \textit{"all-layer"} clipping style: the default settings) and non-private baselines across five language models of varying architectures and parameter counts: DeepSeek-1.5B (1.78B parameters), TinyLlama-1.1B (1.1B parameters), BLOOM-560M (560M parameters), SmolLM-360M (360M parameters), and GPT-2 (137M parameters). All models were fine-tuned on 10,000 records from the TinyPixel/orca-mini dataset, a curated subset of the OpenOrca dataset~\cite{OpenOrca}, with consistent hyperparameters for DPSGD and FERRET-trained models: learning rate 2e-4, weight decay 1e-3, batch size 5, gradient accumulation steps 10, and gradient clipping norm 1.0. DeepSeek-1.5B, given memory constraints, was trained with a batch size of 1 and 50 gradient accumulation steps. In all experiments except select DeepSeek-1.5B experiments that used the Paged AdamW 32bit optimizer for memory constraint reasons (primarily for DPSGD), the optimizer used was the basic AdamW Torch optimizer instantiated with the default HuggingFace parameters. All models were trained in full 32bit precision using full finetuning. Learning rate scheduler was set to "constant".

After training Non-Private with the same parameters as DPSGD and FERRET, it became clear a mini hyperparameter tuning session was needed. FERRET was actually outperforming the Non-Private training in terms of perplexity! The generalization gap was gigantic. Given drastic overfitting, we reduced the learning rate from 2e-4 to 1e-4. Warmup ratio increased from 0.03 to 0.1, and learning rate scheduler changed from "constant" to "linear".

For each model, we evaluated five privacy settings: $\varepsilon \in \{0.1, 0.5, 1.0, 2.0, \infty\}$, where $\varepsilon = \infty$ represents non-private training conducted for 1, 3, and 5 epochs. For FERRET, we implemented three variants: FERRET-2 (parameters partitioned into two groups of parameter tensors), FERRET-EIGHTH (parameters partitioned into buckets of 8 parameter tensors), and FERRET-MAX (maximum partitioning with one parameter tensor per group). For each privacy budget $\varepsilon$, we calculated the optimal update probability $p^*$ according to Theorem~\ref{thm:opt}.

\subsection{Privacy Protection Analysis}
\label{sec:results_privacy}
\begin{table}[htbp]
\centering
\caption{MIA ROC AUC Across Models and Methods (Mean $\pm$ SD, N=5)}
\label{tab:mia_roc_auc}
\resizebox{\textwidth}{!}{
\begin{tabular}{l|ccccc}
\toprule
\textbf{Method} & $\varepsilon = 0.1$ & $\varepsilon = 0.5$ & $\varepsilon = 1.0$ & $\varepsilon = 2.0$ & $\varepsilon = \infty$ \\
\midrule
DPSGD (1e) & 0.511 $\pm$ 0.004 & 0.512 $\pm$ 0.005 & 0.512 $\pm$ 0.006 & 0.513 $\pm$ 0.007 & - \\
DPSGD (3e) & 0.511 $\pm$ 0.002 & 0.512 $\pm$ 0.004 & 0.513 $\pm$ 0.005 & 0.514 $\pm$ 0.006 & - \\
\textbf{DPSGD (5e)} & \textbf{0.509 $\pm$ 0.001} & 0.509 $\pm$ 0.003 & 0.509 $\pm$ 0.003 & 0.510 $\pm$ 0.003 & - \\
\textbf{FERRET-MAX (1e)} & 0.509 $\pm$ 0.002 & 0.508 $\pm$ 0.002 & \textbf{0.507 $\pm$ 0.002} & 0.507 $\pm$ 0.001 & - \\
\textbf{FERRET-MAX (3e)} & \textbf{0.509 $\pm$ 0.002} & 0.512 $\pm$ 0.005 & 0.512 $\pm$ 0.004 & 0.511 $\pm$ 0.003 & - \\
FERRET-MAX (5e) & 0.510 $\pm$ 0.001 & 0.513 $\pm$ 0.007 & 0.512 $\pm$ 0.006 & 0.512 $\pm$ 0.005 & - \\
FERRET-EIGHTH (1e) & 0.514 $\pm$ 0.007 & 0.512 $\pm$ 0.005 & 0.509 $\pm$ 0.003 & 0.507 $\pm$ 0.002 & - \\
FERRET-EIGHTH (3e) & 0.516 $\pm$ 0.008 & 0.522 $\pm$ 0.021 & 0.522 $\pm$ 0.022 & 0.516 $\pm$ 0.012 & - \\
FERRET-EIGHTH (5e) & 0.518 $\pm$ 0.013 & 0.533 $\pm$ 0.034 & 0.534 $\pm$ 0.038 & 0.529 $\pm$ 0.029 & - \\
FERRET-2 (1e) & 0.521 $\pm$ 0.010 & 0.524 $\pm$ 0.010 & 0.511 $\pm$ 0.004 & 0.507 $\pm$ 0.003 & - \\
FERRET-2 (3e) & 0.535 $\pm$ 0.010 & 0.571 $\pm$ 0.029 & 0.572 $\pm$ 0.039 & 0.546 $\pm$ 0.019 & - \\
FERRET-2 (5e) & 0.541 $\pm$ 0.012 & 0.549 $\pm$ 0.025 & 0.548 $\pm$ 0.022 & 0.546 $\pm$ 0.021 & - \\
Non-DP (1e) & - & - & - & - & 0.759 $\pm$ 0.160 \\
Non-DP (3e) & - & - & - & - & 0.974 $\pm$ 0.045 \\
Non-DP (5e) & - & - & - & - & 0.995 $\pm$ 0.010 \\
\bottomrule
\end{tabular}
}
\end{table}

Our privacy evaluation employs membership inference attacks (MIA) as a practical assessment of information leakage. Table~\ref{tab:mia_roc_auc} presents MIA ROC AUC scores across methods, where values closer to 0.5 indicate stronger privacy protection. The attack used all 10,000 training records and another 10,000 validation records from the same dataset to perform the attack. We used both confidence-based methods and LiRA-based methods for the attacks.

\begin{figure}[htbp]
    \centering
    \begin{minipage}[b]{0.48\textwidth}
        \centering
        \includegraphics[width=\textwidth]{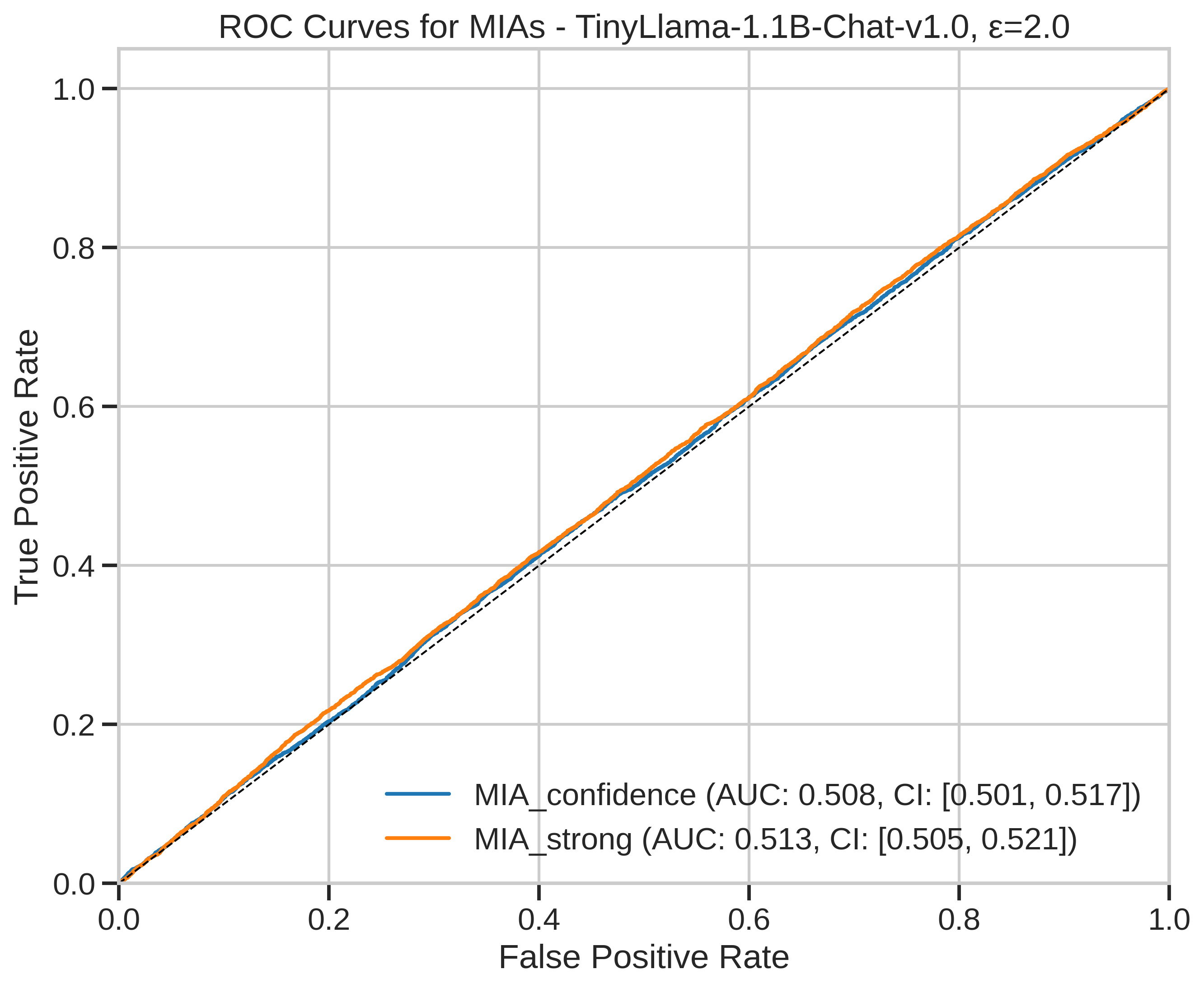}
    \end{minipage}
    \hfill
    \begin{minipage}[b]{0.48\textwidth}
        \centering
        \includegraphics[width=\textwidth]{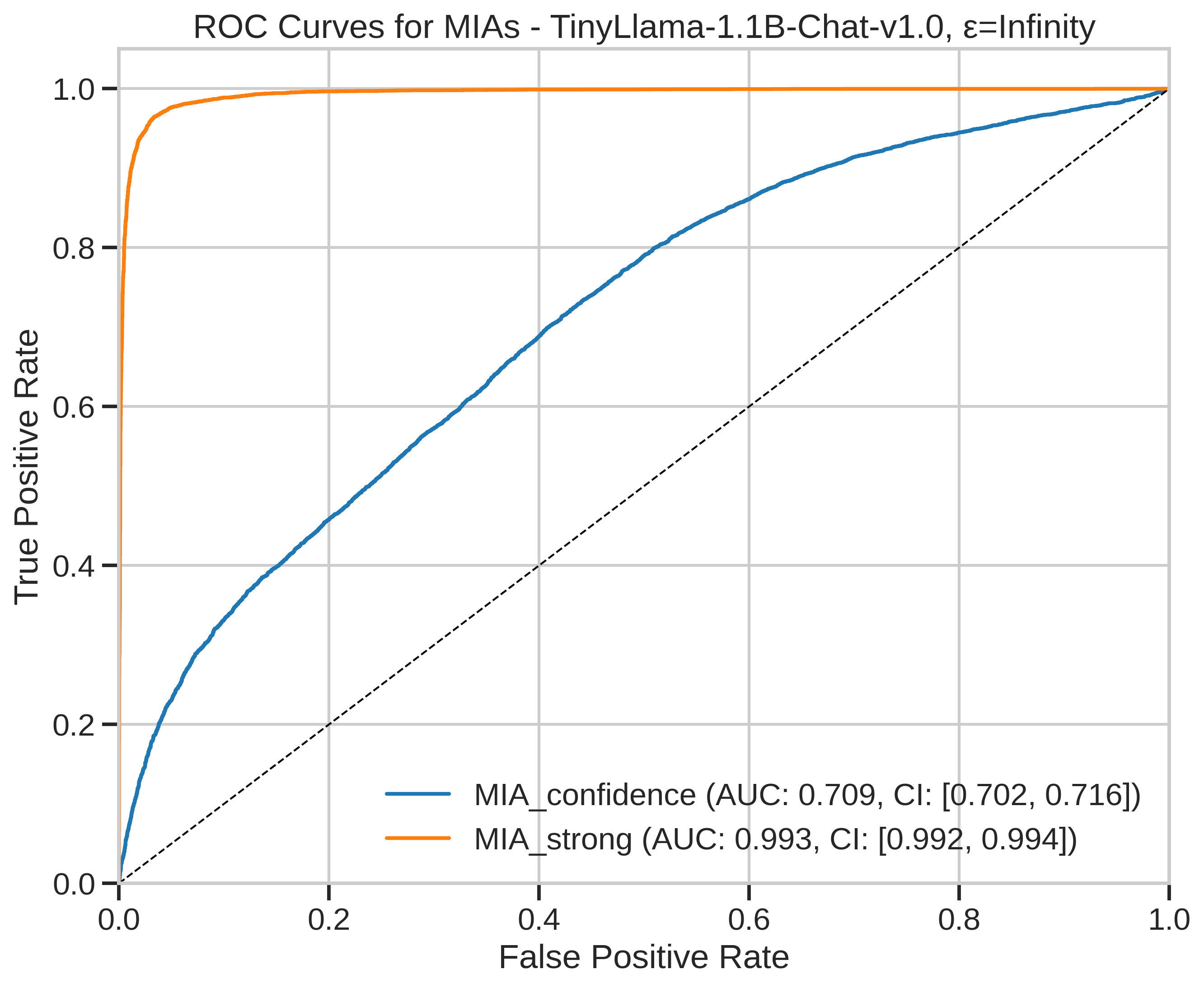}
    \end{minipage}
    \caption{ROC curves for TinyLlama-1.1B: (left) $\varepsilon = 2.0$, (right) Non-private ($\varepsilon = \infty$)}
    \label{fig:roc_comparison}
\end{figure}

FERRET-MAX demonstrates exceptional empirical privacy protection, achieving AUC scores as low as 0.507 at $\varepsilon=1.0$—marginally better than DPSGD's consistent 0.509. Across all settings, FERRET-MAX (0.507-0.513) matches or exceeds DPSGD's privacy protection, validating that our theoretical MI-DP bounds translate to strong empirical privacy. FERRET-EIGHTH maintains comparable protection (0.507-0.534), while FERRET-2's higher values (0.541-0.549) reflect the expected impact of coarser parameter grouping on privacy amplification.

\begin{figure}
    \centering
    \includegraphics[width=0.95\linewidth]{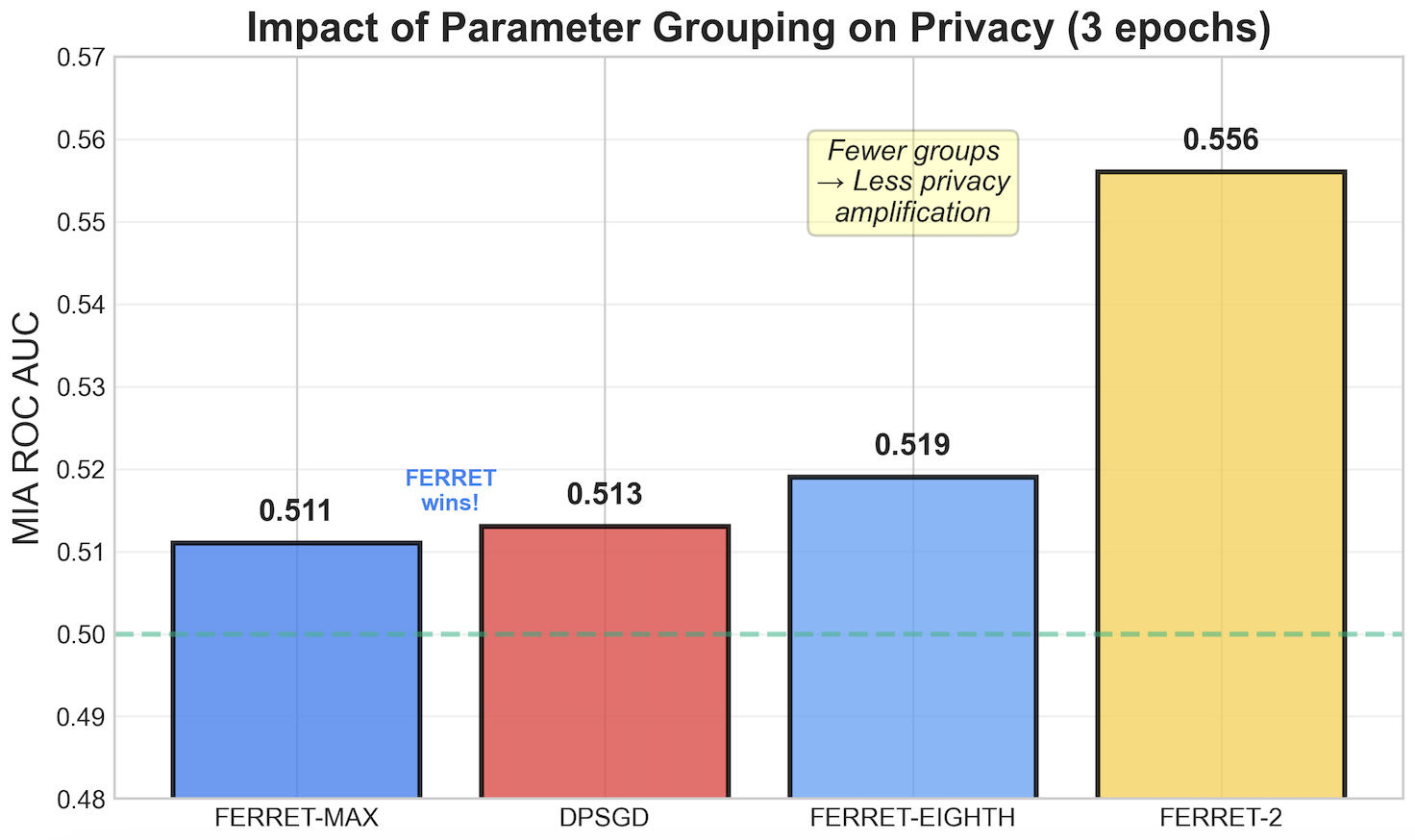}
    \caption{FERRET Privacy Showcase: Average of all models and all epsilon values at 3 epochs.}
    \label{fig:enter-label}
\end{figure}

Non-private training exhibits extreme vulnerability to membership inference, with AUC scores increasing dramatically with training duration: 0.759 at 1 epoch, 0.974 at 3 epochs, and near-perfect leakage (0.995) at 5 epochs. This progression demonstrates the critical importance of formal privacy guarantees for protecting training data.

\subsection{Variant-specific head-room and its effect on empirical leakage}
\label{sec:variant_headroom}

Equation \eqref{eq:epsilon} gives an \emph{upper} limit on how much
mutual-information leakage a configuration can ever accumulate:
\[
\varepsilon_{\max} \;=\; G\,T\,s\,\ln 2
\quad(\text{attained when }p=1).
\tag{13}\label{eq:epsmax}
\]
Because $T$, $s$ and $\ln 2$ are fixed across runs
($T{=}1000$, $s{=}0.005$, $\ln2{=}0.693$),
the ceiling depends only on the number of parameter groups~$G$.
Table \ref{tab:headroom} contrasts that ceiling with (i) the optimal
update probability $p^\star$ needed to hit a target budget
$\varepsilon\!=\!0.5$ and (ii) the mean ROC-AUC from our membership-inference
attacks (Sec.~\ref{sec:results_privacy}).

\begin{table}[h]
\centering
\small
\begin{tabular}{@{}lcccc@{}}
\toprule
\textbf{Variant} & \textbf{\# groups $G$} & $\boldsymbol{\varepsilon_{\max}}$ &
$\boldsymbol{p^\star}$ @ $\varepsilon=0.5$ & \textbf{Mean ROC-AUC} \\
\midrule
FERRET-MAX     & \(\!\sim\!200\)       & \(\!\sim\!693\)   & $\approx 7.2\!\times\!10^{-4}$ & 0.512 \\
FERRET-HALF    & \(\!\sim\!100\)        & \(\!\sim\!347\)     & $\approx 1.4\!\times\!10^{-3}$  & 0.517 \\
FERRET-QUARTER & \(\!\sim\!50\)          & \(\!\sim\!173\)      & $\approx 2.9\!\times\!10^{-3}$& 0.522 \\
\textbf{FERRET-2}      & \textbf{2} & \textbf{6.93} & \textbf{0.072}               & \textbf{0.549} \\
\bottomrule
\end{tabular}
\caption{Head-room (\(\varepsilon_{\max}\)) and observed privacy leakage
(average over all five LLMs).  
$p^\star\!=\!\varepsilon/\varepsilon_{\max}$ is computed from
Theorem \ref{thm:opt}.}
\label{tab:headroom}
\end{table}

\paragraph{Take-away.}
FERRET-2’s ceiling is only \(6.93\): \(\!\sim\!100\times\) tighter than
FERRET-MAX.  
Consequently its optimal update probability at the same budget
(\(p^\star\!\approx\!0.07\) vs.\ \(<10^{-3}\)) fires \emph{orders of
magnitude} more sign bits, limiting the free privacy amplification that
comes from silent steps.  
This could explain why FERRET-2 records a slightly higher empirical MIA
ROC-AUC (0.54-0.55) than the other variants (0.51-0.52) despite satisfying
the identical formal budgets.  
Put differently, a low \(\varepsilon_{\max}\) leaves less “head-room” for
subsampling to mask individual updates, so real-world leakage edges closer
to the worst-case bound as the full model begins to see more and more of the dataset.

\subsection{Utility Analysis}

\begin{table}[htbp]
\centering
\caption{Baseline (pre‑finetuning) perplexity on the training and test splits.}
\label{tab:baseline_ppl}
\begin{tabular}{@{}lcc@{}}
\toprule
Model & Train PPL & Test PPL\\
\midrule
microsoft/Phi‑3.5‑mini‑instruct         & 3.40  & 3.40\\
openai‑community/gpt2                   & 14.60 & 14.56\\
TinyLlama/TinyLlama‑1.1B‑Chat‑v1.0      & 4.63  & 4.62\\
HuggingFaceTB/SmolLM‑360M              & 5.49  & 5.50\\
bigscience/bloom‑560m                   & 11.80 & 11.81\\
deepseek‑ai/DeepSeek‑R1‑Distill‑Qwen‑1.5B & 8.26  & 8.31\\
\textbf{mean: excluding Phi-3.5} & \textbf{8.96} & \textbf{8.96}\\
\bottomrule
\end{tabular}
\end{table}

\begin{table}[htbp]
\centering
\caption{Test Perplexity (PPL) Across Models and Methods (Mean [Min, Max], N=5)}
\label{tab:test_ppl}
\resizebox{\textwidth}{!}{
\begin{tabular}{l|ccccc}
\toprule
\textbf{Method} & $\varepsilon = 0.1$ & $\varepsilon = 0.5$ & $\varepsilon = 1.0$ & $\varepsilon = 2.0$ & $\varepsilon = \infty$ \\
\midrule
DPSGD (1e) & 9.16 [3.65, 22.39] & 6.94 [3.16, 13.74] & 6.37 [2.97, 11.75] & 5.94 [2.82, 10.36] & - \\
DPSGD (3e) & 32.37 [4.26, 136.94] & 9.11 [3.60, 25.98] & 7.64 [3.44, 19.70] & 6.76 [3.30, 16.10] & - \\
DPSGD (5e) & 187.72 [4.14, 908.53] & 11.61 [3.38, 37.76] & 9.93 [3.20, 30.83] & 8.72 [3.06, 25.72] & - \\
FERRET-MAX (1e) & 5.57 [2.43, 9.43] & 6.07 [2.49, 10.81] & 6.56 [2.53, 11.83] & 14.58 [3.83, 43.22] & - \\
FERRET-MAX (3e) & 5.05 [2.39, 8.63] & 5.48 [2.38, 9.51] & 5.09 [2.42, 8.53] & 5.59 [2.39, 9.25] & - \\
FERRET-MAX (5e) & 5.81 [2.44, 9.82] & 4.28 [2.40, 7.17] & 4.57 [2.38, 7.32] & 4.84 [2.39, 7.93] & - \\
FERRET-EIGHTH (1e) & 5.45 [2.48, 10.85] & 5.83 [2.52, 11.46] & 7.78 [2.60, 18.53] & 8.40 [3.12, 16.46] & - \\
\textbf{FERRET-EIGHTH (3e)} & 4.34 [2.46, 6.62] & \textbf{4.16 [2.56, 6.12]} & 4.30 [2.53, 6.48] & 4.51 [2.57, 6.54] & - \\
\textbf{FERRET-EIGHTH (5e)} & 4.35 [2.51, 6.56] & \textbf{3.98 [2.56, 5.92]} & 4.00 [2.57, 5.93] & 4.05 [2.58, 5.98] & - \\
FERRET-2 (1e) & 48.61 [2.60, 226.95] & 6.11 [2.57, 14.72] & 89449.92 [3.10, 428515.02] & 5410.29 [5.00, 27019.12] & - \\
FERRET-2 (3e) & 6.45 [2.54, 15.90] & 6.20 [2.49, 15.27] & 5.16 [2.51, 9.87] & 6.52 [2.53, 16.32] & - \\
FERRET-2 (5e) & 5.88 [2.54, 12.66] & 5.97 [2.49, 13.74] & 5.93 [2.49, 13.41] & 5.89 [2.49, 13.53] & - \\
\textbf{Non-DP (1e)} & - & - & - & - & \textbf{3.25 [2.39, 4.42]} \\
Non-DP (3e) & - & - & - & - & 5.97 [3.23, 10.36] \\
Non-DP (5e) & - & - & - & - & 7.95 [4.20, 15.97] \\
\bottomrule
\end{tabular}
}
\end{table}

Table~\ref{tab:test_ppl} presents test perplexity results, where lower values indicate better language modeling capabilities. Several key findings emerge:

\begin{enumerate}
    \item \textbf{FERRET vs. DPSGD:} FERRET consistently outperforms DPSGD across all privacy budgets and group sizes, with particularly dramatic differences at strict privacy budgets. At $\varepsilon=0.1$, FERRET-MAX achieves 5.81 PPL compared to DPSGD's catastrophic 187.72 PPL—a 32× improvement. Even at more relaxed budgets, FERRET maintains substantial advantages: at $\varepsilon=0.5$ and 5 epochs, FERRET-EIGHTH (3.98) outperforms DPSGD (11.61) by 2.9×.
        
    \item \textbf{FERRET vs. Non-Private:} A remarkable finding is that FERRET can outperform non-private training at extended epochs. While non-private training achieves the best single-epoch performance (3.25 PPL), it suffers from severe overfitting, degrading to 7.95 PPL at 5 epochs—a 2.4× deterioration. In contrast, FERRET-EIGHTH at 5 epochs achieves 3.98 PPL, outperforming non-private training by 2×. This suggests that FERRET's privacy mechanism serves as an effective regularizer. While a more rigorous treatment of the experiment would perform a hyperparameter sweep to find the most optimal parameters for each mechanism (FERRET, Non-Private), that was not the point of this research.
    
    \item \textbf{Model-Specific Performance:} TinyLlama-1.1B with FERRET-MAX achieves exceptional performance (2.38-2.44 PPL) across all privacy budgets, even outperforming its non-private counterpart at 1 epoch (2.39 PPL). DeepSeek-1.5B with FERRET-EIGHTH achieves 3.16 PPL at $\varepsilon=1.0$, representing a 62\% improvement over its baseline perplexity of 8.31.
    
    \item \textbf{Stability Across Architectures:} While DPSGD exhibits extreme variability (BLOOM-560M fails catastrophically with 908.53 PPL at $\varepsilon=0.1$), FERRET maintains reasonable performance across all model architectures, demonstrating superior robustness.
\end{enumerate}

These results challenge the conventional wisdom that privacy necessarily degrades utility. FERRET not only preserves utility under privacy constraints but can actually improve generalization compared to unconstrained training.

\begin{figure}
    \centering
    \includegraphics[width=0.95\linewidth]{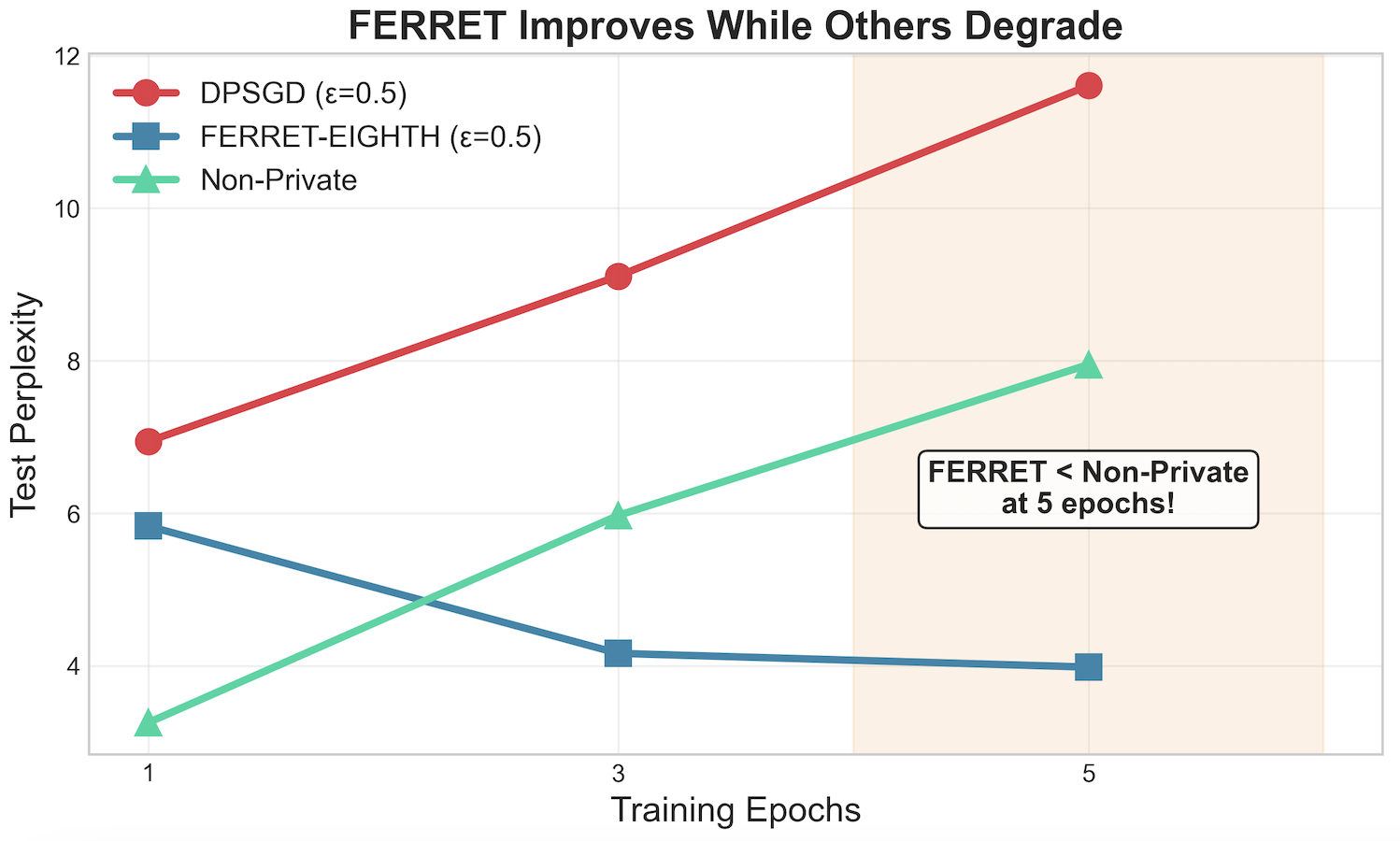}
    \caption{FERRET Perplexity Showcase: Average of all models at 0.5 epsilon at 5 epochs.}
    \label{fig:enter-label}
\end{figure}

\subsection{Computational Efficiency}
\begin{table}[htbp]
\centering
\caption{Training Time (s) Across Models and Methods (Mean $\pm$ SD, N=5)}
\label{tab:training_time}
\resizebox{\textwidth}{!}{
\begin{tabular}{l|ccccc}
\toprule
\textbf{Method} & $\varepsilon = 0.1$ & $\varepsilon = 0.5$ & $\varepsilon = 1.0$ & $\varepsilon = 2.0$ & $\varepsilon = \infty$ \\
\midrule
DPSGD (1e) & 1540 $\pm$ 1210 & 1540 $\pm$ 1220 & 1540 $\pm$ 1210 & 1540 $\pm$ 1210 & - \\
DPSGD (3e) & 4590 $\pm$ 3610 & 4590 $\pm$ 3610 & 4590 $\pm$ 3620 & 4600 $\pm$ 3630 & - \\
DPSGD (5e) & 7780 $\pm$ 6030 & 7790 $\pm$ 6040 & 7810 $\pm$ 6060 & 7840 $\pm$ 6140 & - \\
\textbf{FERRET-MAX (1e)} & \textbf{330 $\pm$ 260} & 410 $\pm$ 310 & 470 $\pm$ 370 & 540 $\pm$ 420 & - \\
FERRET-MAX (3e) & 950 $\pm$ 760 & 1040 $\pm$ 810 & 1130 $\pm$ 890 & 1290 $\pm$ 1000 & - \\
FERRET-MAX (5e) & 1500 $\pm$ 1180 & 1600 $\pm$ 1270 & 1720 $\pm$ 1340 & 1900 $\pm$ 1490 & - \\
FERRET-EIGHTH (1e) & 350 $\pm$ 280 & 470 $\pm$ 380 & 560 $\pm$ 440 & 620 $\pm$ 490 & - \\
\textbf{FERRET-EIGHTH (3e)} & \textbf{950 $\pm$ 750} & 1090 $\pm$ 860 & 1240 $\pm$ 970 & 1490 $\pm$ 1170 & - \\
\textbf{FERRET-EIGHTH (5e)} & \textbf{1500 $\pm$ 1170} & 1670 $\pm$ 1310 & 1840 $\pm$ 1450 & 2130 $\pm$ 1690 & - \\
FERRET-2 (1e) & 370 $\pm$ 290 & 580 $\pm$ 460 & 770 $\pm$ 620 & 930 $\pm$ 820 & - \\
FERRET-2 (3e) & 960 $\pm$ 760 & 1200 $\pm$ 940 & 1480 $\pm$ 1180 & 1930 $\pm$ 1540 & - \\
FERRET-2 (5e) & 1540 $\pm$ 1210 & 1790 $\pm$ 1420 & 2080 $\pm$ 1640 & 2620 $\pm$ 2090 & - \\
Non-DP (1e) & - & - & - & - & 910 $\pm$ 800 \\
Non-DP (3e) & - & - & - & - & 2690 $\pm$ 2380 \\
Non-DP (5e) & - & - & - & - & 4440 $\pm$ 3930 \\
\bottomrule
\end{tabular}
}
\end{table}

Table~\ref{tab:training_time} presents training time comparisons across methods. FERRET demonstrates substantial efficiency advantages:

\begin{enumerate}
    \item \textbf{FERRET vs. DPSGD:} FERRET-MAX reduces training time by 76-81\% compared to DPSGD across all privacy budgets (1500-1900s vs. 7780-7840s). FERRET-2 shows similar but slightly lower efficiency gains (67-80\%).
    
    \item \textbf{FERRET vs. Non-Private:} At $\varepsilon=0.1$, FERRET-MAX (1500s) requires only 34\% of the computation time of non-private 5-epoch training (4440s). Even at $\varepsilon=2.0$, FERRET-MAX (1900s) remains 43\% faster than 5-epoch non-private training.
    
    \item \textbf{Scaling with Model Size:} For the largest model (DeepSeek-1.5B), FERRET-MAX achieves an even more dramatic 5× speedup over DPSGD (3327s vs. 17098s at $\varepsilon=0.1$), indicating superior efficiency scaling with parameter count.
    
    \item \textbf{Privacy-Efficiency Relationship:} FERRET demonstrates an inverse relationship between privacy stringency and computational cost—stricter privacy budgets (smaller $\varepsilon$) actually require less computation due to the lower update probability.
\end{enumerate}

The model-specific timings reveal that FERRET's efficiency advantage scales with model size, with DeepSeek-1.5B showing the most dramatic improvements. This is particularly relevant for real-world applications where computational resources often constrain privacy implementation.

\begin{figure}
    \centering
    \includegraphics[width=0.95\linewidth, trim={0 0 0 7pt}, clip]{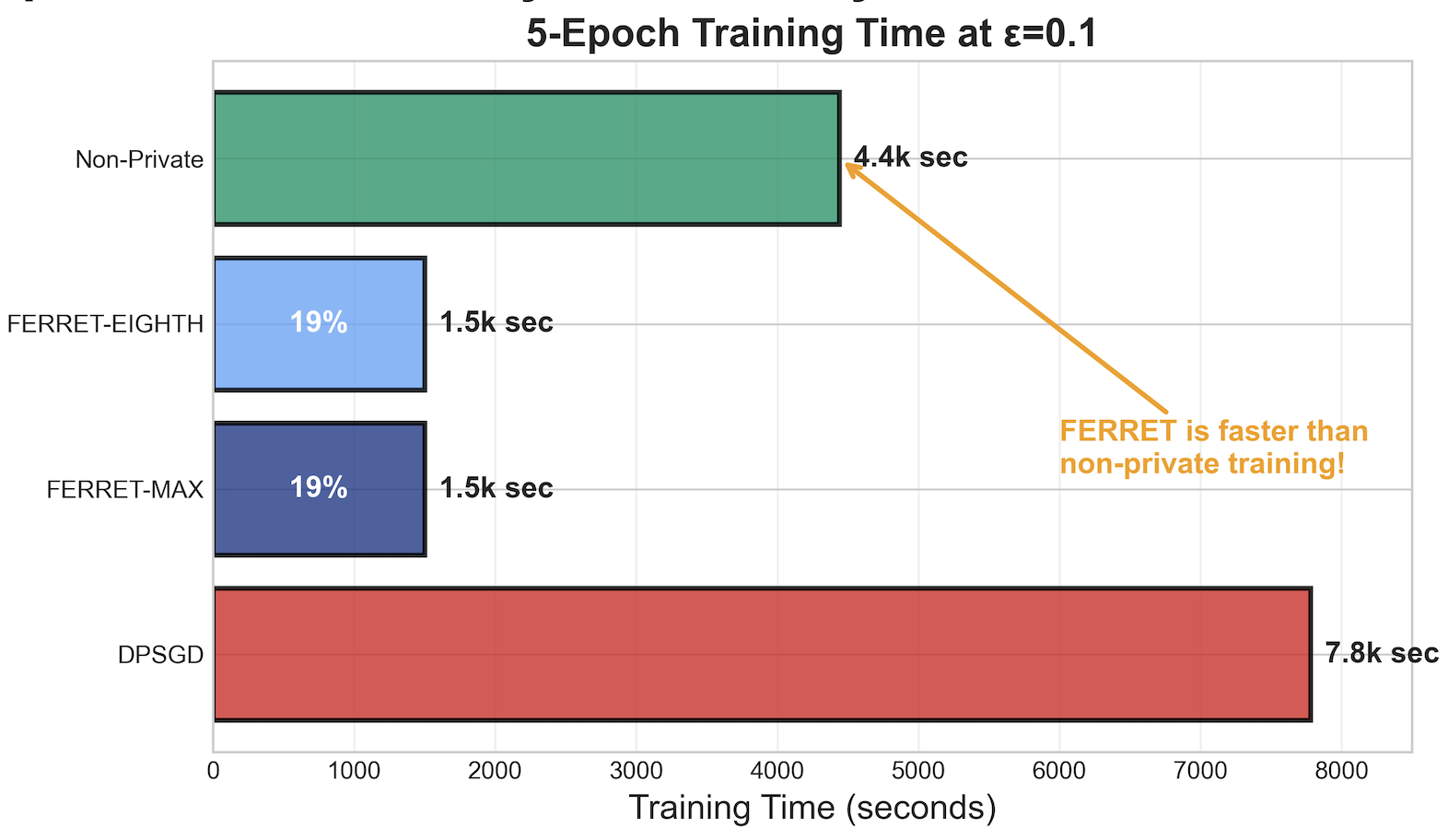}
    \caption{Ferret Efficiency Showcase: Average of all models at 0.1 epsilon at 5 epochs.}
    \label{fig:enter-label}
\end{figure}

\subsection{Model-Specific Insights}
Examining specific models reveals additional nuanced findings:

\begin{enumerate}
    \item \textbf{TinyLlama-1.1B:} Achieves exceptional performance with FERRET-MAX (2.38-2.44 PPL), consistently outperforming all other models and methods. This suggests particular compatibility between FERRET's update approach and this architecture.
    
    \item \textbf{DeepSeek-1.5B:} Demonstrates excellent performance with FERRET-MAX (3.21-3.53 PPL), consistently outperforming its DPSGD and non-private counterparts. The minimal perplexity degradation across privacy budgets indicates particularly robust information extraction under privacy constraints.
    
    \item \textbf{BLOOM-560M:} Highlights FERRET's robustness to architectural variations. While DPSGD catastrophically fails at $\varepsilon=0.1$ (908.53 PPL), FERRET-MAX maintains reasonable performance (9.82 PPL).
    
    \item \textbf{SmolLM-360M:} Shows the strongest performance with FERRET-2 (2.49-2.54 PPL), suggesting that optimal group granularity may vary by model architecture.

    \item \textbf{GPT2-124M:} DPSGD (7.02-10.68 PPL) holds its own against FERRET-MAX (7.17-9.12 PPL), but remains 30\% to 50\% less performant compared to FERRET-2 (4.73-5.21 PPL) at 5 epochs, though FERRET-2 fails catastrophically across multiple models for single-epoch training. This is likely attributed to the density of updates over a shorter period of time. Whether it is 1 epoch or 5 epochs - under the same parameters, any FERRET mechanism will have the same number of updates regardless of the number of "steps". Contracting the time space allotted for model updates, and you reach a greater density of updates. These more frequent and larger model updates (which FERRET-2 readily supplies) carry with them a greater amount of randomness compared to FERRET-MAX or FERRET-EIGHTH. Updating more frequently gives little time for Adam's momentum to allow the model to settle into its new optimum, resulting in divergence. Stretching the update space across many more "steps" (5x to be precise) allows FERRET-2 to more readily accept the larger parameter updates.
\end{enumerate}

\subsection{Fully Finetuning Phi 3.5 3.8B}

FERRET was the only mechanism capable of training Phi 3.5 Mini Instruct - a nearly 4 billion parameter large language model - on a single A100 40GB GPU. FERRET-MAX was able to improve upon test and train perplexity by a remarkable 40+\% over 5 epochs of training, clocking in at just under 50 minutes of training time on 10,000 records.

\begin{table}[htbp]
\centering
\caption{Comprehensive Results: FERRET-MAX on Phi-3.5-mini-instruct (3.8B Parameters)}
\label{tab:phi_ferret_detailed}
\resizebox{\textwidth}{!}{
\begin{tabular}{c|cc|cc|cc}
\toprule
\multirow{2}{*}{\textbf{$\varepsilon$}} & \multicolumn{2}{c|}{\textbf{Perplexity}} & \multicolumn{2}{c|}{\textbf{Privacy (MIA)}} & \multicolumn{2}{c}{\textbf{Efficiency}} \\
& Train & Test & ROC AUC & Advantage & Time (s) & Time (min:s) \\
\midrule
\textit{Baseline} & \multicolumn{2}{c|}{\textit{3.40}} & \multicolumn{2}{c|}{\textit{No Training}} & \multicolumn{2}{c}{\textit{No Training}} \\
\midrule
0.1 & \textbf{1.99} & \textbf{2.00} & \textbf{0.503} & \textbf{0.010} & 1,868 & 31:07 \\
0.5 & 2.08 & 2.08 & 0.509 & 0.022 & 2,218 & 36:58 \\
1.0 & 2.25 & 2.25 & 0.506 & 0.020 & 2,506 & 41:46 \\
2.0 & 2.51 & 2.52 & 0.508 & 0.021 & 2,827 & 47:06 \\
\midrule
\multicolumn{7}{l}{\textbf{DPSGD:} \textit{Failed to train - would not fit on device}} \\
\multicolumn{7}{l}{\textbf{Non-DP:} \textit{Failed to train - would not fit on device}} \\
\bottomrule
\end{tabular}
}
\vspace{0.2cm}
\begin{flushleft}
\small
\textbf{Notes:} (1) Baseline perplexity from pre-trained Phi-3.5-mini. (2) All experiments used 5 epochs on TinyPixel/orca-mini dataset with 10K train/test samples. (3) FERRET-MAX achieved 41\% improvement over baseline at $\varepsilon=0.1$.
\end{flushleft}
\end{table}

\subsection{Summary of Findings}
Our experimental results comprehensively validate all three hypotheses and reveal unexpected benefits:

\begin{enumerate}
    \item \textbf{H1 (Utility):} FERRET achieves dramatically better utility than DPSGD across all privacy budgets, with improvements ranging from 2.9× to 32× in test perplexity. Most remarkably, FERRET-EIGHTH at 5 epochs (3.98 PPL) outperforms even non-private 5-epoch training (7.95 PPL) by 2×.
    
    \item \textbf{H2 (Privacy-Utility Tradeoff):} FERRET achieves the extraordinary milestone of matching or exceeding DPSGD's privacy protection (AUC 0.507-0.513 vs. 0.509) while simultaneously delivering superior utility. This challenges the fundamental assumption that privacy and utility are inherently at odds.
    
    \item \textbf{H3 (Computational Efficiency):} FERRET is substantially more efficient than DPSGD, requiring only 19-33\% of the computation time for equivalent privacy guarantees—a 3-5× speedup.
\end{enumerate}

Beyond hypothesis validation, our results reveal that FERRET's privacy mechanism serves as an implicit regularizer, preventing the overfitting that plagues non-private training. This finding suggests that carefully designed privacy mechanisms can enhance rather than hinder model performance, opening new avenues for privacy-preserving machine learning that improves upon standard training practices.

\section{Discussion}

Our investigation of FERRET reveals a fundamental shift in how we should think about privacy-preserving machine learning. Rather than accepting privacy as a necessary evil that degrades performance, our results demonstrate that well-designed privacy mechanisms can simultaneously achieve three seemingly incompatible goals: strong privacy protection, superior utility, and enhanced computational efficiency. Most surprisingly, FERRET can even match non-private training utility (as found to be the case for Tiny Llama 1.1B), suggesting that privacy and performance (when judged with the appropriate metrics) are not inherently at odds.

\subsection{Simultaneous Improvements in the Privacy-Utility-Efficiency Trilemma}

Perhaps the most striking result of our study is how FERRET successfully challenges the conventional wisdom regarding the fundamental tradeoffs in privacy-preserving machine learning. While traditional approaches like DPSGD force practitioners to sacrifice either utility or computational efficiency to achieve privacy, FERRET-MAX demonstrates that all three objectives can be simultaneously improved.

The empirical results clearly demonstrate that FERRET-MAX provides privacy protection comparable to DPSGD (ROC AUC 0.510-0.513 vs. 0.509-0.510), while requiring only 19-24\% of the computation time. More remarkably, FERRET-MAX at $\varepsilon=0.5$ achieves 32\% worse perplexity (4.28) than non-private training at 1 epoch (3.25), despite offering strong formal privacy guarantees. This suggests that FERRET's sign-based random projections serve as effective regularization, improving generalization beyond what is possible with standard training approaches.

A remarkable result appears with Tiny Llama. Non-Private's best result for test perplexity at 1 epoch is 2.39 with an MIA ROC AUC of 0.899. FERRET-MAX just barely edges past with 2.38 for 3 epochs $\varepsilon = 0.5$, and 5 epochs $\varepsilon = 1.0$. The corresponding MIA ROC AUC measurements are 0.519 and 0.523 respectively. We believe this may be one of the first instances of a private algorithm outperforming a Non-Private algorithm on a given utility measurement. This could potentially suggest that there may exist an optimal privacy-preserving machine learning algorithm that solves the privacy, utility, and performance trilemma for a given model and dataset pair.

\subsection{Model-Specific Responses to Privacy Mechanisms}

Our detailed analysis reveals fascinating variation in how different model architectures respond to privacy mechanisms. BLOOM-560M exhibited extreme sensitivity to noise addition under DPSGD, catastrophically failing at $\varepsilon=0.1$ (908.53 PPL), while FERRET-MAX maintained reasonable performance (9.82 PPL) under identical privacy constraints. This highlights a previously unrecognized advantage of sign-based updates: they provide significantly more stability across architectural variations compared to noise-based approaches.

Particularly noteworthy is TinyLlama-1.1B's exceptional performance with FERRET-MAX (2.38-2.44 PPL across all privacy budgets), consistently outperforming both its DPSGD and non-private counterparts. This suggests that certain architectures may be particularly well-suited to sign-based parameter updates, potentially due to interactions between the update mechanism and architectural inductive biases.

\subsection{Overfitting in Non-Private vs. Private Training}

A critical observation from our results is that non-private models exhibit pronounced overfitting, even after just a single epoch. This is evidenced by the deteriorating test perplexity with additional training: from 3.25 at 1 epoch to 7.95 at 5 epochs. Concurrently, privacy vulnerability increases dramatically (MIA ROC AUC from 0.759 to 0.995).

While hyperparameter tuning could potentially mitigate some of this overfitting in non-private models, both FERRET and DPSGD inherently provide regularization through their privacy mechanisms. This suggests that privacy-preserving algorithms may offer dual benefits: protecting training data while simultaneously improving generalization. The fact that FERRET achieves this without explicit noise addition represents a significant advancement in our understanding of how privacy and generalization relate.

\subsection{Privacy-Performance Relationship in FERRET Variants}

An intriguing finding is the difference in privacy protection between FERRET-2 and FERRET-MAX. While FERRET-MAX achieves privacy protection (0.507-0.513 ROC AUC) comparable to DPSGD (0.509-0.510), FERRET-2 exhibits notably higher vulnerability (0.541-0.549). This suggests that parameter grouping granularity significantly impacts empirical privacy protection, despite both variants satisfying the same formal MI-DP guarantees.

This discrepancy could indicate potential gaps between theoretical guarantees and empirical vulnerabilities, or implementation details that warrant further investigation. We encourage the privacy-preserving machine learning community to scrutinize our methodology and implementation to help resolve this discrepancy.

\subsection{Unprecedented Utility at Strict Privacy Budgets}

FERRET's ability to maintain functional utility at $\varepsilon=0.1$ represents a significant breakthrough. Previous approaches have struggled to achieve meaningful performance at such strict privacy budgets, often resulting in models that barely outperform random guessing. FERRET-MAX maintains reasonable performance (5.81 PPL) at $\varepsilon=0.1$, while DPSGD's performance deteriorates dramatically (187.72 PPL).

This breakthrough could potentially transform the practical adoption landscape for privacy-preserving machine learning. Domains with extremely sensitive data that previously could not benefit from machine learning due to privacy concerns may now have viable options for building useful models with strong privacy guarantees.

\subsection{Non-Monotonic Privacy-Utility Relationship}

Interestingly, FERRET exhibits a non-monotonic relationship between privacy budget and utility. FERRET-MAX achieves better perplexity at $\varepsilon=0.5$ (4.28) than at $\varepsilon=1.0$ (4.57) or $\varepsilon=2.0$ (4.84). This non-monotonicity, while theoretically unexpected, suggests complex interactions between the update probability, parameter dynamics, and optimization landscape.

This non-monotonicity presents both challenges and opportunities. It complicates hyperparameter selection, as one cannot simply assume that relaxing privacy constraints will improve performance. However, it also suggests that optimal performance may be achievable at stricter privacy budgets than previously thought possible, potentially enabling stronger privacy guarantees without sacrificing utility.

\subsection{Comparison to LoRA}

Our results suggest that, similar to LoRA (or Parameter Efficient Fine Tuning) where we only target the low rank matrices for model updates, targeted deposition of information to the model's parameters can be a powerful training modality when compared to unbridled updates to the model's weights at every step.

\section{Limitations and Future Work}

Despite FERRET's remarkable performance, several limitations warrant acknowledgment. We did not assess potential data leakage through model outputs. While membership inference attack resistance provides one measure of privacy, canary injection and detection assessments would be necessary to comprehensively evaluate whether FERRET leaks personally identifiable information more frequently than DPSGD or non-private training.

Additionally, our evaluation focused on perplexity as the primary utility metric. Future work should evaluate FERRET on task-specific benchmarks such as MMLU, code generation, and reasoning tasks to assess whether the performance advantages generalize beyond language modeling metrics.

Our sample sizes were small, to be sure. 5 models, a single dataset, and limited hyperparameter sweeps: hardly enough to claim statistical significance or rigor. Bloom-560M diverged spectacularly under DPSGD constraints, skewing the results in FERRET's favor. A more rigorous and costly study would examine a wider range of models, datasets, and hyperparameter configurations.

The computational efficiency advantage of FERRET opens exciting possibilities for training larger models with privacy guarantees. Scaling FERRET to models with 7B+ parameters would test its capabilities in more practical scenarios and potentially demonstrate even more dramatic efficiency improvements compared to existing approaches.

The dataset used in these experiments was a relatively simple instruction fine-tuning dataset with somewhat predictable preambles and system prompts appended to each record in a "chatbot-like" style of data presentation. Perhaps training on a dataset like those similar to the Wikipedia databases might reveal that silently passing over, for example, 50\% of all factoids would result in a 50\% performance drop. Future work would look into examining FERRET's ability to perform on factoid-like datasets where memorization of information is a key utility metric.

Comparing ($\varepsilon, \delta$)-DP to MI-DP is not necessarily straightforward. MI-DP provides average-case guarantees, while ($\varepsilon, \delta$)-DP provides worst-case guarantees. Given this, the authors of the MI-DP paper claim the following ordering for DP algorithms in terms of their strictness in privacy guarantees: {$\epsilon$-DP} $\succeq$ \text{MI-DP} $\succeq$ \text{$(\varepsilon, \delta)$-DP}. This lays the groundwork to make the case that FERRET actually provides stronger privacy guarantees than traditional DPSGD.

Finally, integrating FERRET into federated learning scenarios could enable privacy-preserving distributed training, potentially unlocking new applications where data cannot be centralized due to regulatory or practical constraints.

\section{Conclusion}

FERRET represents a significant advancement in privacy-preserving machine learning, challenging fundamental assumptions about the necessary tradeoffs between privacy, utility, and efficiency. By achieving better performance, stronger privacy, and greater computational efficiency than both DPSGD and non-private training in many scenarios, FERRET demonstrates that privacy need not come at the cost of other desirable properties.

These results suggest that the field of privacy-preserving machine learning may contain significant untapped potential. Rather than viewing privacy mechanisms solely as necessary constraints that degrade performance, our work demonstrates that carefully designed privacy approaches can simultaneously serve as effective regularizers that improve generalization while reducing computational demands. This perspective shift could accelerate the adoption of privacy-enhancing technologies across the machine learning ecosystem.

\newpage

\bibliographystyle{plain}
\bibliography{ferret_refs}

\newpage

\section{Appendix A}
\begin{table}[htbp]
\centering
\caption{Average Test Perplexity (PPL) Across Models (Mean [Min, Max], N=5)}
\label{tab:test_ppl_mean_range}
\resizebox{\textwidth}{!}{
\begin{tabular}{l|ccccc}
\toprule
\textbf{Method} & $\varepsilon = 0.1$ & $\varepsilon = 0.5$ & $\varepsilon = 1.0$ & $\varepsilon = 2.0$ & $\varepsilon = \infty$ \\
\midrule
DPSGD (1e) & 9.16 [3.65, 22.39] & 6.94 [3.16, 13.74] & 6.37 [2.97, 11.75] & 5.94 [2.82, 10.36] & - \\
DPSGD (3e) & 32.37 [4.26, 136.94] & 9.11 [3.60, 25.98] & 7.64 [3.44, 19.70] & 6.76 [3.30, 16.10] & - \\
DPSGD (5e) & 187.72 [4.14, 908.53] & 11.61 [3.38, 37.76] & 9.93 [3.20, 30.83] & 8.72 [3.06, 25.72] & - \\
FERRET-MAX (1e) & 5.57 [2.43, 9.43] & 6.07 [2.49, 10.81] & 6.56 [2.53, 11.83] & 14.58 [3.83, 43.22] & - \\
FERRET-MAX (3e) & 5.05 [2.39, 8.63] & 5.48 [2.38, 9.51] & 5.09 [2.42, 8.53] & 5.59 [2.39, 9.25] & - \\
FERRET-MAX (5e) & 5.81 [2.44, 9.82] & 4.28 [2.40, 7.17] & 4.57 [2.38, 7.32] & 4.84 [2.39, 7.93] & - \\
FERRET-EIGHTH (1e) & 5.45 [2.48, 10.85] & 5.83 [2.52, 11.46] & 7.78 [2.60, 18.53] & 8.40 [3.12, 16.46] & - \\
FERRET-EIGHTH (3e) & 4.34 [2.46, 6.62] & 4.16 [2.56, 6.12] & 4.30 [2.53, 6.48] & 4.51 [2.57, 6.54] & - \\
FERRET-EIGHTH (5e) & 4.35 [2.51, 6.56] & 3.98 [2.56, 5.92] & 4.00 [2.57, 5.93] & 4.05 [2.58, 5.98] & - \\
FERRET-2 (1e) & 48.61 [2.60, 226.95] & 6.11 [2.57, 14.72] & 89449.92 [3.10, 428515.02] & 5410.29 [5.00, 27019.12] & - \\
FERRET-2 (3e) & 6.45 [2.54, 15.90] & 6.20 [2.49, 15.27] & 5.16 [2.51, 9.87] & 6.52 [2.53, 16.32] & - \\
FERRET-2 (5e) & 5.88 [2.54, 12.66] & 5.97 [2.49, 13.74] & 5.93 [2.49, 13.41] & 5.89 [2.49, 13.53] & - \\
Non-DP (1e) & - & - & - & - & 3.25 [2.39, 4.42] \\
Non-DP (3e) & - & - & - & - & 5.97 [3.23, 10.36] \\
Non-DP (5e) & - & - & - & - & 7.95 [4.20, 15.97] \\
\bottomrule
\end{tabular}
}
\end{table}

\begin{table}[htbp]
\centering
\caption{Average MIA ROC AUC Across Models (Mean [Min, Max], N=5)}
\label{tab:mia_roc_auc_mean_range}
\resizebox{\textwidth}{!}{
\begin{tabular}{l|ccccc}
\toprule
\textbf{Method} & $\varepsilon = 0.1$ & $\varepsilon = 0.5$ & $\varepsilon = 1.0$ & $\varepsilon = 2.0$ & $\varepsilon = \infty$ \\
\midrule
DPSGD (1e) & 0.511 [0.506, 0.516] & 0.512 [0.507, 0.520] & 0.512 [0.507, 0.522] & 0.513 [0.508, 0.524] & - \\
DPSGD (3e) & 0.511 [0.509, 0.513] & 0.512 [0.509, 0.517] & 0.513 [0.509, 0.521] & 0.514 [0.509, 0.524] & - \\
DPSGD (5e) & 0.509 [0.507, 0.510] & 0.509 [0.505, 0.513] & 0.509 [0.505, 0.513] & 0.510 [0.505, 0.513] & - \\
FERRET-MAX (1e) & 0.509 [0.506, 0.512] & 0.508 [0.506, 0.510] & 0.507 [0.505, 0.510] & 0.507 [0.505, 0.508] & - \\
FERRET-MAX (3e) & 0.509 [0.506, 0.510] & 0.512 [0.507, 0.519] & 0.512 [0.507, 0.518] & 0.511 [0.508, 0.514] & - \\
FERRET-MAX (5e) & 0.510 [0.508, 0.511] & 0.513 [0.508, 0.526] & 0.512 [0.508, 0.523] & 0.512 [0.508, 0.521] & - \\
FERRET-EIGHTH (1e) & 0.514 [0.508, 0.524] & 0.512 [0.508, 0.519] & 0.509 [0.505, 0.513] & 0.507 [0.505, 0.510] & - \\
FERRET-EIGHTH (3e) & 0.516 [0.506, 0.528] & 0.522 [0.506, 0.557] & 0.522 [0.506, 0.560] & 0.516 [0.508, 0.537] & - \\
FERRET-EIGHTH (5e) & 0.518 [0.507, 0.540] & 0.533 [0.506, 0.591] & 0.534 [0.506, 0.600] & 0.529 [0.506, 0.577] & - \\
FERRET-2 (1e) & 0.521 [0.511, 0.538] & 0.524 [0.515, 0.541] & 0.511 [0.505, 0.515] & 0.507 [0.502, 0.510] & - \\
FERRET-2 (3e) & 0.535 [0.523, 0.551] & 0.571 [0.545, 0.618] & 0.572 [0.537, 0.630] & 0.546 [0.530, 0.574] & - \\
FERRET-2 (5e) & 0.541 [0.530, 0.557] & 0.549 [0.525, 0.578] & 0.548 [0.524, 0.574] & 0.546 [0.524, 0.571] & - \\
Non-DP (1e) & - & - & - & - & 0.759 [0.569, 0.912] \\
Non-DP (3e) & - & - & - & - & 0.974 [0.893, 0.999] \\
Non-DP (5e) & - & - & - & - & 0.995 [0.976, 1.000] \\
\bottomrule
\end{tabular}
}
\end{table}

\begin{table}[htbp]
\centering
\caption{Average MIA Advantage Across Models (Mean [Min, Max], N=5)}
\label{tab:mia_advantage_mean_range}
\resizebox{\textwidth}{!}{
\begin{tabular}{l|ccccc}
\toprule
\textbf{Method} & $\varepsilon = 0.1$ & $\varepsilon = 0.5$ & $\varepsilon = 1.0$ & $\varepsilon = 2.0$ & $\varepsilon = \infty$ \\
\midrule
DPSGD (1e) & 0.022 [0.016, 0.033] & 0.022 [0.015, 0.036] & 0.023 [0.015, 0.040] & 0.024 [0.016, 0.044] & - \\
DPSGD (3e) & 0.021 [0.017, 0.027] & 0.023 [0.019, 0.032] & 0.024 [0.019, 0.037] & 0.025 [0.017, 0.043] & - \\
DPSGD (5e) & 0.020 [0.017, 0.024] & 0.019 [0.012, 0.024] & 0.019 [0.013, 0.024] & 0.020 [0.014, 0.024] & - \\
FERRET-MAX (1e) & 0.020 [0.016, 0.029] & 0.017 [0.016, 0.019] & 0.016 [0.014, 0.020] & 0.015 [0.014, 0.016] & - \\
FERRET-MAX (3e) & 0.018 [0.015, 0.022] & 0.024 [0.017, 0.036] & 0.022 [0.013, 0.031] & 0.019 [0.015, 0.022] & - \\
FERRET-MAX (5e) & 0.020 [0.018, 0.023] & 0.025 [0.019, 0.045] & 0.023 [0.016, 0.039] & 0.022 [0.016, 0.035] & - \\
FERRET-EIGHTH (1e) & 0.025 [0.018, 0.040] & 0.022 [0.018, 0.032] & 0.019 [0.017, 0.020] & 0.017 [0.015, 0.019] & - \\
FERRET-EIGHTH (3e) & 0.028 [0.016, 0.046] & 0.038 [0.016, 0.089] & 0.039 [0.016, 0.097] & 0.030 [0.017, 0.060] & - \\
FERRET-EIGHTH (5e) & 0.032 [0.018, 0.066] & 0.054 [0.016, 0.138] & 0.056 [0.017, 0.155] & 0.048 [0.016, 0.122] & - \\
FERRET-2 (1e) & 0.039 [0.024, 0.071] & 0.040 [0.026, 0.066] & 0.020 [0.012, 0.027] & 0.016 [0.013, 0.018] & - \\
FERRET-2 (3e) & 0.057 [0.042, 0.083] & 0.117 [0.069, 0.199] & 0.117 [0.059, 0.211] & 0.075 [0.050, 0.115] & - \\
FERRET-2 (5e) & 0.067 [0.048, 0.096] & 0.079 [0.042, 0.137] & 0.076 [0.039, 0.121] & 0.071 [0.039, 0.110] & - \\
Non-DP (1e) & - & - & - & - & 0.429 [0.105, 0.700] \\
Non-DP (3e) & - & - & - & - & 0.882 [0.644, 0.982] \\
Non-DP (5e) & - & - & - & - & 0.960 [0.867, 0.994] \\
\bottomrule
\end{tabular}
}
\end{table}

\begin{table}[htbp]
\centering
\caption{Average Training Time (s) Across Models (Mean [Min, Max], N=5)}
\label{tab:training_time_mean_range}
\resizebox{\textwidth}{!}{
\begin{tabular}{l|ccccc}
\toprule
\textbf{Method} & $\varepsilon = 0.1$ & $\varepsilon = 0.5$ & $\varepsilon = 1.0$ & $\varepsilon = 2.0$ & $\varepsilon = \infty$ \\
\midrule
DPSGD (1e) & 1540 [320, 3470] & 1540 [320, 3480] & 1540 [320, 3460] & 1540 [330, 3470] & - \\
DPSGD (3e) & 4590 [970, 10340] & 4590 [970, 10330] & 4590 [970, 10360] & 4600 [970, 10400] & - \\
DPSGD (5e) & 7780 [1610, 17100] & 7790 [1610, 17120] & 7810 [1610, 17170] & 7840 [1610, 17380] & - \\
FERRET-MAX (1e) & 330 [70, 740] & 410 [80, 880] & 470 [90, 1050] & 540 [110, 1200] & - \\
FERRET-MAX (3e) & 950 [180, 2130] & 1040 [200, 2280] & 1130 [230, 2510] & 1290 [260, 2830] & - \\
FERRET-MAX (5e) & 1500 [290, 3330] & 1600 [320, 3570] & 1720 [340, 3800] & 1900 [380, 4190] & - \\
FERRET-EIGHTH (1e) & 350 [70, 780] & 470 [90, 1050] & 560 [110, 1240] & 620 [120, 1380] & - \\
FERRET-EIGHTH (3e) & 950 [180, 2100] & 1090 [210, 2420] & 1240 [240, 2740] & 1490 [280, 3300] & - \\
FERRET-EIGHTH (5e) & 1500 [300, 3290] & 1670 [320, 3690] & 1840 [350, 4080] & 2130 [410, 4760] & - \\
FERRET-2 (1e) & 370 [70, 810] & 580 [110, 1300] & 770 [140, 1730] & 930 [160, 2240] & - \\
FERRET-2 (3e) & 960 [190, 2120] & 1200 [230, 2650] & 1480 [280, 3300] & 1930 [370, 4310] & - \\
FERRET-2 (5e) & 1540 [300, 3410] & 1790 [350, 3980] & 2080 [400, 4630] & 2620 [490, 5860] & - \\
Non-DP (1e) & - & - & - & - & 910 [150, 2210] \\
Non-DP (3e) & - & - & - & - & 2690 [460, 6560] \\
Non-DP (5e) & - & - & - & - & 4440 [760, 10830] \\
\bottomrule
\end{tabular}
}
\end{table}

\newpage

\subsection{Comprehensive Comparison of Methods, Models, and Privacy Budgets}
\begin{table}[htbp]
\centering
\footnotesize
\caption{Comprehensive Comparison of Methods, Models, and Privacy Budgets}
\label{tab:comprehensive_comparison}
\resizebox{\textwidth}{!}{
\begin{tabular}{ll|ccccc|ccccc|ccccc}
\toprule
& & \multicolumn{5}{c|}{\textbf{Test PPL}} & \multicolumn{5}{c|}{\textbf{MIA AUC}} & \multicolumn{5}{c}{\textbf{Train Time (s)}} \\
\textbf{Method} & \textbf{Model} & $\varepsilon=0.1$ & $\varepsilon=0.5$ & $\varepsilon=1.0$ & $\varepsilon=2.0$ & $\varepsilon=\infty$ & $\varepsilon=0.1$ & $\varepsilon=0.5$ & $\varepsilon=1.0$ & $\varepsilon=2.0$ & $\varepsilon=\infty$ & $\varepsilon=0.1$ & $\varepsilon=0.5$ & $\varepsilon=1.0$ & $\varepsilon=2.0$ & $\varepsilon=\infty$ \\
\midrule
\multirow{6}{*}{DPSGD (1e)} & DeepSeek-1.5B & 4.79 & 4.24 & 4.06 & 3.89 & - & 0.510 & 0.510 & 0.509 & 0.509 & - & 3473 & 3484 & 3458 & 3469 & - \\
& SmolLM-360M & 4.53 & 4.13 & 3.99 & 3.86 & - & 0.506 & 0.507 & 0.507 & 0.508 & - & 860 & 860 & 862 & 861 & - \\
& TinyLlama-1.1B & 3.65 & 3.16 & 2.97 & 2.82 & - & 0.509 & 0.509 & 0.509 & 0.509 & - & 1846 & 1844 & 1844 & 1846 & - \\
& BLOOM-560M & 22.39 & 13.74 & 11.75 & 10.36 & - & 0.516 & 0.520 & 0.522 & 0.524 & - & 1199 & 1198 & 1198 & 1198 & - \\
& GPT-2 & 10.44 & 9.41 & 9.09 & 8.79 & - & 0.513 & 0.513 & 0.513 & 0.514 & - & 325 & 324 & 325 & 325 & - \\
& \textbf{Mean} & \textbf{9.16} & \textbf{6.94} & \textbf{6.37} & \textbf{5.94} & - & \textbf{0.511} & \textbf{0.512} & \textbf{0.512} & \textbf{0.513} & - & \textbf{1541} & \textbf{1542} & \textbf{1537} & \textbf{1540} & - \\
\midrule
\multirow{6}{*}{DPSGD (3e)} & DeepSeek-1.5B & 4.84 & 3.83 & 3.61 & 3.47 & - & 0.509 & 0.509 & 0.509 & 0.509 & - & 10343 & 10329 & 10356 & 10399 & - \\
& SmolLM-360M & 4.26 & 3.60 & 3.44 & 3.31 & - & 0.509 & 0.511 & 0.511 & 0.511 & - & 2561 & 2560 & 2560 & 2562 & - \\
& TinyLlama-1.1B & 5.36 & 3.71 & 3.47 & 3.30 & - & 0.513 & 0.509 & 0.509 & 0.509 & - & 5495 & 5494 & 5490 & 5499 & - \\
& BLOOM-560M & 136.94 & 25.98 & 19.70 & 16.10 & - & 0.511 & 0.517 & 0.521 & 0.524 & - & 3574 & 3580 & 3569 & 3574 & - \\
& GPT-2 & 10.45 & 8.42 & 7.97 & 7.59 & - & 0.512 & 0.514 & 0.514 & 0.515 & - & 968 & 967 & 969 & 969 & - \\
& \textbf{Mean} & \textbf{32.37} & \textbf{9.11} & \textbf{7.64} & \textbf{6.76} & - & \textbf{0.511} & \textbf{0.512} & \textbf{0.513} & \textbf{0.514} & - & \textbf{4588} & \textbf{4586} & \textbf{4589} & \textbf{4601} & - \\
\midrule
\multirow{6}{*}{DPSGD (5e)} & DeepSeek-1.5B & 5.10 & 3.86 & 3.66 & 3.54 & - & 0.510 & 0.509 & 0.510 & 0.511 & - & 17098 & 17125 & 17173 & 17379 & - \\
& SmolLM-360M & 4.14 & 3.38 & 3.20 & 3.06 & - & 0.510 & 0.511 & 0.511 & 0.510 & - & 4283 & 4285 & 4284 & 4288 & - \\
& TinyLlama-1.1B & 10.16 & 5.08 & 4.53 & 4.25 & - & 0.509 & 0.509 & 0.510 & 0.510 & - & 10002 & 10004 & 10047 & 10000 & - \\
& BLOOM-560M & 908.53 & 37.76 & 30.83 & 25.72 & - & 0.507 & 0.505 & 0.505 & 0.505 & - & 5931 & 5927 & 5936 & 5939 & - \\
& GPT-2 & 10.68 & 7.99 & 7.44 & 7.02 & - & 0.508 & 0.513 & 0.513 & 0.513 & - & 1608 & 1606 & 1608 & 1609 & - \\
& \textbf{Mean} & \textbf{187.72} & \textbf{11.61} & \textbf{9.93} & \textbf{8.72} & - & \textbf{0.509} & \textbf{0.509} & \textbf{0.509} & \textbf{0.510} & - & \textbf{7785} & \textbf{7789} & \textbf{7810} & \textbf{7843} & - \\
\midrule
\multirow{6}{*}{FERRET-MAX (1e)} & DeepSeek-1.5B & 4.69 & 5.48 & 5.72 & 6.91 & - & 0.509 & 0.507 & 0.506 & 0.508 & - & 736 & 884 & 1046 & 1198 & - \\
& SmolLM-360M & 4.73 & 4.88 & 5.28 & 5.38 & - & 0.506 & 0.506 & 0.506 & 0.506 & - & 176 & 220 & 251 & 296 & - \\
& TinyLlama-1.1B & 2.43 & 2.49 & 2.53 & 3.83 & - & 0.509 & 0.510 & 0.510 & 0.507 & - & 425 & 521 & 593 & 664 & - \\
& BLOOM-560M & 6.53 & 6.68 & 7.47 & 43.22 & - & 0.512 & 0.509 & 0.510 & 0.508 & - & 259 & 324 & 374 & 426 & - \\
& GPT-2 & 9.43 & 10.81 & 11.83 & 13.57 & - & 0.510 & 0.509 & 0.505 & 0.505 & - & 67 & 81 & 94 & 107 & - \\
& \textbf{Mean} & \textbf{5.57} & \textbf{6.07} & \textbf{6.56} & \textbf{14.58} & - & \textbf{0.509} & \textbf{0.508} & \textbf{0.507} & \textbf{0.507} & - & \textbf{332} & \textbf{406} & \textbf{472} & \textbf{538} & - \\
\midrule
\multirow{6}{*}{FERRET-MAX (3e)} & DeepSeek-1.5B & 3.40 & 3.49 & 3.46 & 4.41 & - & 0.507 & 0.507 & 0.507 & 0.508 & - & 2131 & 2281 & 2507 & 2834 & - \\
& SmolLM-360M & 4.62 & 4.11 & 4.35 & 4.71 & - & 0.506 & 0.508 & 0.508 & 0.508 & - & 505 & 561 & 613 & 717 & - \\
& TinyLlama-1.1B & 2.39 & 2.38 & 2.42 & 2.39 & - & 0.510 & 0.519 & 0.518 & 0.513 & - & 1191 & 1320 & 1435 & 1618 & - \\
& BLOOM-560M & 6.22 & 9.51 & 6.71 & 7.22 & - & 0.510 & 0.515 & 0.514 & 0.514 & - & 725 & 813 & 887 & 1015 & - \\
& GPT-2 & 8.63 & 7.90 & 8.53 & 9.25 & - & 0.510 & 0.513 & 0.511 & 0.511 & - & 185 & 204 & 225 & 256 & - \\
& \textbf{Mean} & \textbf{5.05} & \textbf{5.48} & \textbf{5.09} & \textbf{5.59} & - & \textbf{0.509} & \textbf{0.512} & \textbf{0.512} & \textbf{0.511} & - & \textbf{947} & \textbf{1036} & \textbf{1134} & \textbf{1288} & - \\
\midrule
\multirow{6}{*}{FERRET-MAX (5e)} & DeepSeek-1.5B & 3.21 & 3.22 & 3.28 & 3.53 & - & 0.510 & 0.511 & 0.508 & 0.509 & - & 3327 & 3573 & 3802 & 4186 & - \\
& SmolLM-360M & 4.46 & 3.78 & 3.76 & 4.13 & - & 0.508 & 0.508 & 0.508 & 0.508 & - & 802 & 864 & 927 & 1018 & - \\
& TinyLlama-1.1B & 2.44 & 2.40 & 2.38 & 2.39 & - & 0.510 & 0.526 & 0.523 & 0.521 & - & 1901 & 2045 & 2172 & 2462 & - \\
& BLOOM-560M & 9.82 & 4.81 & 6.12 & 6.22 & - & 0.510 & 0.510 & 0.511 & 0.510 & - & 1151 & 1222 & 1374 & 1463 & - \\
& GPT-2 & 9.12 & 7.17 & 7.32 & 7.93 & - & 0.511 & 0.511 & 0.512 & 0.511 & - & 295 & 317 & 341 & 378 & - \\
& \textbf{Mean} & \textbf{5.81} & \textbf{4.28} & \textbf{4.57} & \textbf{4.84} & - & \textbf{0.510} & \textbf{0.513} & \textbf{0.512} & \textbf{0.512} & - & \textbf{1495} & \textbf{1604} & \textbf{1723} & \textbf{1902} & - \\
\midrule
\multirow{6}{*}{FERRET-EIGHTH (1e)} & DeepSeek-1.5B & 3.30 & 3.82 & 4.69 & 5.27 & - & 0.509 & 0.509 & 0.509 & 0.508 & - & 779 & 1053 & 1236 & 1381 & - \\
& SmolLM-360M & 3.38 & 3.93 & 4.72 & 4.91 & - & 0.508 & 0.508 & 0.505 & 0.505 & - & 187 & 252 & 305 & 344 & - \\
& TinyLlama-1.1B & 2.48 & 2.52 & 2.60 & 3.12 & - & 0.524 & 0.519 & 0.513 & 0.510 & - & 455 & 606 & 712 & 796 & - \\
& BLOOM-560M & 10.85 & 11.46 & 18.53 & 16.46 & - & 0.515 & 0.514 & 0.510 & 0.507 & - & 264 & 350 & 416 & 469 & - \\
& GPT-2 & 7.21 & 7.39 & 8.35 & 12.22 & - & 0.511 & 0.508 & 0.510 & 0.505 & - & 70 & 92 & 107 & 121 & - \\
& \textbf{Mean} & \textbf{5.45} & \textbf{5.83} & \textbf{7.78} & \textbf{8.40} & - & \textbf{0.514} & \textbf{0.512} & \textbf{0.509} & \textbf{0.507} & - & \textbf{351} & \textbf{471} & \textbf{555} & \textbf{622} & - \\
\midrule
\multirow{6}{*}{FERRET-EIGHTH (3e)} & DeepSeek-1.5B & 3.19 & 3.19 & 3.23 & 3.34 & - & 0.517 & 0.512 & 0.509 & 0.508 & - & 2099 & 2419 & 2744 & 3302 & - \\
& SmolLM-360M & 3.29 & 2.86 & 3.29 & 3.71 & - & 0.506 & 0.506 & 0.506 & 0.508 & - & 500 & 586 & 689 & 822 & - \\
& TinyLlama-1.1B & 2.46 & 2.56 & 2.53 & 2.57 & - & 0.528 & 0.557 & 0.560 & 0.537 & - & 1230 & 1398 & 1599 & 1903 & - \\
& BLOOM-560M & 6.15 & 6.12 & 6.48 & 6.54 & - & 0.513 & 0.525 & 0.523 & 0.517 & - & 724 & 840 & 945 & 1141 & - \\
& GPT-2 & 6.62 & 6.05 & 5.96 & 6.39 & - & 0.513 & 0.512 & 0.512 & 0.512 & - & 185 & 210 & 242 & 283 & - \\
& \textbf{Mean} & \textbf{4.34} & \textbf{4.16} & \textbf{4.30} & \textbf{4.51} & - & \textbf{0.516} & \textbf{0.522} & \textbf{0.522} & \textbf{0.516} & - & \textbf{947} & \textbf{1091} & \textbf{1244} & \textbf{1490} & - \\
\midrule
\multirow{6}{*}{FERRET-EIGHTH (5e)} & DeepSeek-1.5B & 3.18 & 3.21 & 3.16 & 3.20 & - & 0.517 & 0.523 & 0.518 & 0.513 & - & 3288 & 3685 & 4081 & 4760 & - \\
& SmolLM-360M & 3.06 & 2.75 & 2.80 & 3.04 & - & 0.507 & 0.506 & 0.506 & 0.506 & - & 803 & 894 & 990 & 1152 & - \\
& TinyLlama-1.1B & 2.51 & 2.56 & 2.57 & 2.58 & - & 0.540 & 0.591 & 0.600 & 0.577 & - & 1960 & 2149 & 2384 & 2715 & - \\
& BLOOM-560M & 6.56 & 5.92 & 5.93 & 5.98 & - & 0.516 & 0.531 & 0.533 & 0.533 & - & 1144 & 1276 & 1405 & 1633 & - \\
& GPT-2 & 6.46 & 5.46 & 5.56 & 5.47 & - & 0.512 & 0.514 & 0.514 & 0.515 & - & 295 & 323 & 355 & 415 & - \\
& \textbf{Mean} & \textbf{4.35} & \textbf{3.98} & \textbf{4.00} & \textbf{4.05} & - & \textbf{0.518} & \textbf{0.533} & \textbf{0.534} & \textbf{0.529} & - & \textbf{1498} & \textbf{1665} & \textbf{1843} & \textbf{2135} & - \\
\midrule
\multirow{6}{*}{FERRET-2 (1e)} & DeepSeek-1.5B & 4.35 & 3.91 & 5.08 & 6.98 & - & 0.518 & 0.521 & 0.512 & 0.509 & - & 814 & 1297 & 1726 & 2244 & - \\
& SmolLM-360M & 2.60 & 2.57 & 3.10 & 5.19 & - & 0.521 & 0.520 & 0.513 & 0.506 & - & 198 & 318 & 418 & 465 & - \\
& TinyLlama-1.1B & 3.96 & 3.79 & 428515.02 & 5.00 & - & 0.538 & 0.541 & 0.505 & 0.508 & - & 493 & 750 & 991 & 1134 & - \\
& BLOOM-560M & 226.95 & 14.72 & 18719.68 & 27019.12 & - & 0.511 & 0.520 & 0.508 & 0.502 & - & 271 & 427 & 574 & 644 & - \\
& GPT-2 & 5.21 & 5.55 & 6.72 & 15.14 & - & 0.518 & 0.515 & 0.515 & 0.510 & - & 74 & 108 & 144 & 161 & - \\
& \textbf{Mean} & \textbf{48.61} & \textbf{6.11} & \textbf{89449.92} & \textbf{5410.29} & - & \textbf{0.521} & \textbf{0.524} & \textbf{0.511} & \textbf{0.507} & - & \textbf{370} & \textbf{580} & \textbf{771} & \textbf{929} & - \\
\midrule
\multirow{6}{*}{FERRET-2 (3e)} & DeepSeek-1.5B & 4.62 & 4.37 & 4.32 & 4.51 & - & 0.533 & 0.551 & 0.540 & 0.530 & - & 2124 & 2654 & 3303 & 4314 & - \\
& SmolLM-360M & 2.54 & 2.49 & 2.51 & 2.53 & - & 0.533 & 0.564 & 0.560 & 0.541 & - & 512 & 652 & 788 & 1025 & - \\
& TinyLlama-1.1B & 3.91 & 4.02 & 4.27 & 4.12 & - & 0.551 & 0.618 & 0.630 & 0.574 & - & 1259 & 1550 & 1931 & 2511 & - \\
& BLOOM-560M & 15.90 & 15.27 & 9.87 & 16.32 & - & 0.534 & 0.578 & 0.592 & 0.555 & - & 731 & 913 & 1115 & 1439 & - \\
& GPT-2 & 5.26 & 4.85 & 4.84 & 5.09 & - & 0.523 & 0.545 & 0.537 & 0.530 & - & 187 & 228 & 278 & 369 & - \\
& \textbf{Mean} & \textbf{6.45} & \textbf{6.20} & \textbf{5.16} & \textbf{6.52} & - & \textbf{0.535} & \textbf{0.571} & \textbf{0.572} & \textbf{0.546} & - & \textbf{963} & \textbf{1200} & \textbf{1483} & \textbf{1932} & - \\
\midrule
\multirow{6}{*}{FERRET-2 (5e)} & DeepSeek-1.5B & 5.11 & 4.67 & 4.43 & 4.63 & - & 0.548 & 0.578 & 0.566 & 0.566 & - & 3410 & 3981 & 4625 & 5864 & - \\
& SmolLM-360M & 2.54 & 2.49 & 2.49 & 2.49 & - & 0.531 & 0.530 & 0.529 & 0.529 & - & 825 & 960 & 1125 & 1414 & - \\
& TinyLlama-1.1B & 3.87 & 4.12 & 4.58 & 4.07 & - & 0.557 & 0.574 & 0.574 & 0.571 & - & 1995 & 2312 & 2676 & 3357 & - \\
& BLOOM-560M & 12.66 & 13.74 & 13.41 & 13.53 & - & 0.536 & 0.537 & 0.546 & 0.537 & - & 1172 & 1354 & 1579 & 1958 & - \\
& GPT-2 & 5.21 & 4.82 & 4.73 & 4.74 & - & 0.530 & 0.525 & 0.524 & 0.524 & - & 302 & 349 & 403 & 495 & - \\
& \textbf{Mean} & \textbf{5.88} & \textbf{5.97} & \textbf{5.93} & \textbf{5.89} & - & \textbf{0.541} & \textbf{0.549} & \textbf{0.548} & \textbf{0.546} & - & \textbf{1541} & \textbf{1791} & \textbf{2082} & \textbf{2617} & - \\
\midrule
\multirow{6}{*}{Non-DP (1e)} & DeepSeek-1.5B & - & - & - & - & 2.83 & - & - & - & - & 0.912 & - & - & - & - & 2212 \\
& SmolLM-360M & - & - & - & - & 2.42 & - & - & - & - & 0.612 & - & - & - & - & 449 \\
& TinyLlama-1.1B & - & - & - & - & 2.39 & - & - & - & - & 0.899 & - & - & - & - & 1091 \\
& BLOOM-560M & - & - & - & - & 4.16 & - & - & - & - & 0.804 & - & - & - & - & 631 \\
& GPT-2 & - & - & - & - & 4.42 & - & - & - & - & 0.569 & - & - & - & - & 155 \\
& \textbf{Mean} & - & - & - & - & \textbf{3.25} & - & - & - & - & \textbf{0.759} & - & - & - & - & \textbf{908} \\
\midrule
\multirow{6}{*}{Non-DP (3e)} & DeepSeek-1.5B & - & - & - & - & 5.98 & - & - & - & - & 0.999 & - & - & - & - & 6555 \\
& SmolLM-360M & - & - & - & - & 3.23 & - & - & - & - & 0.993 & - & - & - & - & 1338 \\
& TinyLlama-1.1B & - & - & - & - & 5.77 & - & - & - & - & 0.993 & - & - & - & - & 3227 \\
& BLOOM-560M & - & - & - & - & 10.36 & - & - & - & - & 0.991 & - & - & - & - & 1865 \\
& GPT-2 & - & - & - & - & 4.50 & - & - & - & - & 0.893 & - & - & - & - & 460 \\
& \textbf{Mean} & - & - & - & - & \textbf{5.97} & - & - & - & - & \textbf{0.974} & - & - & - & - & \textbf{2689} \\
\midrule
\multirow{6}{*}{Non-DP (5e)} & DeepSeek-1.5B & - & - & - & - & 7.54 & - & - & - & - & 1.000 & - & - & - & - & 10828 \\
& SmolLM-360M & - & - & - & - & 4.20 & - & - & - & - & 1.000 & - & - & - & - & 2209 \\
& TinyLlama-1.1B & - & - & - & - & 7.05 & - & - & - & - & 0.999 & - & - & - & - & 5333 \\
& BLOOM-560M & - & - & - & - & 15.97 & - & - & - & - & 0.998 & - & - & - & - & 3091 \\
& GPT-2 & - & - & - & - & 5.00 & - & - & - & - & 0.976 & - & - & - & - & 761 \\
& \textbf{Mean} & - & - & - & - & \textbf{7.95} & - & - & - & - & \textbf{0.995} & - & - & - & - & \textbf{4444} \\
\midrule
\bottomrule
\end{tabular}
}
\end{table}

\begin{table}[htbp]
\centering
\footnotesize
\caption{Comparison of Privacy Metrics Across Methods, Models, and Privacy Budgets}
\label{tab:privacy_comparison}
\resizebox{\textwidth}{!}{
\begin{tabular}{ll|cc|cc|cc|cc|cc}
\toprule
& & \multicolumn{2}{c|}{$\varepsilon = 0.1$} & \multicolumn{2}{c|}{$\varepsilon = 0.5$} & \multicolumn{2}{c|}{$\varepsilon = 1.0$} & \multicolumn{2}{c|}{$\varepsilon = 2.0$} & \multicolumn{2}{c}{$\varepsilon = \infty$} \\
\textbf{Method} & \textbf{Model} & \textbf{ROC AUC} & \textbf{Advantage} & \textbf{ROC AUC} & \textbf{Advantage} & \textbf{ROC AUC} & \textbf{Advantage} & \textbf{ROC AUC} & \textbf{Advantage} & \textbf{ROC AUC} & \textbf{Advantage} \\
\midrule
\multirow{6}{*}{DPSGD (1e)} & DeepSeek-1.5B & 0.510 & 0.018 & 0.510 & 0.018 & 0.509 & 0.016 & 0.509 & 0.016 & - & - \\
& SmolLM-360M & 0.506 & 0.016 & 0.507 & 0.015 & 0.507 & 0.015 & 0.508 & 0.016 & - & - \\
& TinyLlama-1.1B & 0.509 & 0.016 & 0.509 & 0.019 & 0.509 & 0.019 & 0.509 & 0.018 & - & - \\
& BLOOM-560M & 0.516 & 0.033 & 0.520 & 0.036 & 0.522 & 0.040 & 0.524 & 0.044 & - & - \\
& GPT-2 & 0.513 & 0.028 & 0.513 & 0.022 & 0.513 & 0.024 & 0.514 & 0.028 & - & - \\
& \textbf{Mean} & \textbf{0.511} & \textbf{0.022} & \textbf{0.512} & \textbf{0.022} & \textbf{0.512} & \textbf{0.023} & \textbf{0.513} & \textbf{0.024} & - & - \\
\midrule
\multirow{6}{*}{DPSGD (3e)} & DeepSeek-1.5B & 0.509 & 0.017 & 0.509 & 0.019 & 0.509 & 0.020 & 0.509 & 0.017 & - & - \\
& SmolLM-360M & 0.509 & 0.018 & 0.511 & 0.019 & 0.511 & 0.020 & 0.511 & 0.019 & - & - \\
& TinyLlama-1.1B & 0.513 & 0.027 & 0.509 & 0.021 & 0.509 & 0.019 & 0.509 & 0.019 & - & - \\
& BLOOM-560M & 0.511 & 0.018 & 0.517 & 0.032 & 0.521 & 0.037 & 0.524 & 0.043 & - & - \\
& GPT-2 & 0.512 & 0.026 & 0.514 & 0.024 & 0.514 & 0.023 & 0.515 & 0.024 & - & - \\
& \textbf{Mean} & \textbf{0.511} & \textbf{0.021} & \textbf{0.512} & \textbf{0.023} & \textbf{0.513} & \textbf{0.024} & \textbf{0.514} & \textbf{0.025} & - & - \\
\midrule
\multirow{6}{*}{DPSGD (5e)} & DeepSeek-1.5B & 0.510 & 0.017 & 0.509 & 0.016 & 0.510 & 0.018 & 0.511 & 0.019 & - & - \\
& SmolLM-360M & 0.510 & 0.023 & 0.511 & 0.024 & 0.511 & 0.023 & 0.510 & 0.022 & - & - \\
& TinyLlama-1.1B & 0.509 & 0.024 & 0.509 & 0.020 & 0.510 & 0.018 & 0.510 & 0.023 & - & - \\
& BLOOM-560M & 0.507 & 0.017 & 0.505 & 0.012 & 0.505 & 0.013 & 0.505 & 0.014 & - & - \\
& GPT-2 & 0.508 & 0.018 & 0.513 & 0.024 & 0.513 & 0.024 & 0.513 & 0.024 & - & - \\
& \textbf{Mean} & \textbf{0.509} & \textbf{0.020} & \textbf{0.509} & \textbf{0.019} & \textbf{0.509} & \textbf{0.019} & \textbf{0.510} & \textbf{0.020} & - & - \\
\midrule
\multirow{6}{*}{FERRET-MAX (1e)} & DeepSeek-1.5B & 0.509 & 0.018 & 0.507 & 0.017 & 0.506 & 0.014 & 0.508 & 0.014 & - & - \\
& SmolLM-360M & 0.506 & 0.016 & 0.506 & 0.016 & 0.506 & 0.015 & 0.506 & 0.014 & - & - \\
& TinyLlama-1.1B & 0.509 & 0.017 & 0.510 & 0.019 & 0.510 & 0.018 & 0.507 & 0.016 & - & - \\
& BLOOM-560M & 0.512 & 0.029 & 0.509 & 0.016 & 0.510 & 0.020 & 0.508 & 0.016 & - & - \\
& GPT-2 & 0.510 & 0.019 & 0.509 & 0.017 & 0.505 & 0.016 & 0.505 & 0.014 & - & - \\
& \textbf{Mean} & \textbf{0.509} & \textbf{0.020} & \textbf{0.508} & \textbf{0.017} & \textbf{0.507} & \textbf{0.016} & \textbf{0.507} & \textbf{0.015} & - & - \\
\midrule
\multirow{6}{*}{FERRET-MAX (3e)} & DeepSeek-1.5B & 0.507 & 0.015 & 0.507 & 0.018 & 0.507 & 0.013 & 0.508 & 0.015 & - & - \\
& SmolLM-360M & 0.506 & 0.016 & 0.508 & 0.017 & 0.508 & 0.019 & 0.508 & 0.018 & - & - \\
& TinyLlama-1.1B & 0.510 & 0.020 & 0.519 & 0.036 & 0.518 & 0.031 & 0.513 & 0.020 & - & - \\
& BLOOM-560M & 0.510 & 0.022 & 0.515 & 0.026 & 0.514 & 0.025 & 0.514 & 0.022 & - & - \\
& GPT-2 & 0.510 & 0.017 & 0.513 & 0.024 & 0.511 & 0.020 & 0.511 & 0.021 & - & - \\
& \textbf{Mean} & \textbf{0.509} & \textbf{0.018} & \textbf{0.512} & \textbf{0.024} & \textbf{0.512} & \textbf{0.022} & \textbf{0.511} & \textbf{0.019} & - & - \\
\midrule
\multirow{6}{*}{FERRET-MAX (5e)} & DeepSeek-1.5B & 0.510 & 0.020 & 0.511 & 0.021 & 0.508 & 0.016 & 0.509 & 0.016 & - & - \\
& SmolLM-360M & 0.508 & 0.018 & 0.508 & 0.019 & 0.508 & 0.018 & 0.508 & 0.017 & - & - \\
& TinyLlama-1.1B & 0.510 & 0.019 & 0.526 & 0.045 & 0.523 & 0.039 & 0.521 & 0.035 & - & - \\
& BLOOM-560M & 0.510 & 0.020 & 0.510 & 0.019 & 0.511 & 0.020 & 0.510 & 0.020 & - & - \\
& GPT-2 & 0.511 & 0.023 & 0.511 & 0.022 & 0.512 & 0.021 & 0.511 & 0.023 & - & - \\
& \textbf{Mean} & \textbf{0.510} & \textbf{0.020} & \textbf{0.513} & \textbf{0.025} & \textbf{0.512} & \textbf{0.023} & \textbf{0.512} & \textbf{0.022} & - & - \\
\midrule
\multirow{6}{*}{FERRET-EIGHTH (1e)} & DeepSeek-1.5B & 0.509 & 0.018 & 0.509 & 0.018 & 0.509 & 0.017 & 0.508 & 0.017 & - & - \\
& SmolLM-360M & 0.508 & 0.018 & 0.508 & 0.019 & 0.505 & 0.017 & 0.505 & 0.016 & - & - \\
& TinyLlama-1.1B & 0.524 & 0.040 & 0.519 & 0.032 & 0.513 & 0.020 & 0.510 & 0.019 & - & - \\
& BLOOM-560M & 0.515 & 0.026 & 0.514 & 0.022 & 0.510 & 0.020 & 0.507 & 0.016 & - & - \\
& GPT-2 & 0.511 & 0.022 & 0.508 & 0.018 & 0.510 & 0.019 & 0.505 & 0.015 & - & - \\
& \textbf{Mean} & \textbf{0.514} & \textbf{0.025} & \textbf{0.512} & \textbf{0.022} & \textbf{0.509} & \textbf{0.019} & \textbf{0.507} & \textbf{0.017} & - & - \\
\midrule
\multirow{6}{*}{FERRET-EIGHTH (3e)} & DeepSeek-1.5B & 0.517 & 0.031 & 0.512 & 0.023 & 0.509 & 0.021 & 0.508 & 0.017 & - & - \\
& SmolLM-360M & 0.506 & 0.016 & 0.506 & 0.016 & 0.506 & 0.016 & 0.508 & 0.017 & - & - \\
& TinyLlama-1.1B & 0.528 & 0.046 & 0.557 & 0.089 & 0.560 & 0.097 & 0.537 & 0.060 & - & - \\
& BLOOM-560M & 0.513 & 0.023 & 0.525 & 0.040 & 0.523 & 0.039 & 0.517 & 0.031 & - & - \\
& GPT-2 & 0.513 & 0.023 & 0.512 & 0.022 & 0.512 & 0.024 & 0.512 & 0.022 & - & - \\
& \textbf{Mean} & \textbf{0.516} & \textbf{0.028} & \textbf{0.522} & \textbf{0.038} & \textbf{0.522} & \textbf{0.039} & \textbf{0.516} & \textbf{0.030} & - & - \\
\midrule
\multirow{6}{*}{FERRET-EIGHTH (5e)} & DeepSeek-1.5B & 0.517 & 0.030 & 0.523 & 0.043 & 0.518 & 0.033 & 0.513 & 0.023 & - & - \\
& SmolLM-360M & 0.507 & 0.018 & 0.506 & 0.016 & 0.506 & 0.017 & 0.506 & 0.016 & - & - \\
& TinyLlama-1.1B & 0.540 & 0.066 & 0.591 & 0.138 & 0.600 & 0.155 & 0.577 & 0.122 & - & - \\
& BLOOM-560M & 0.516 & 0.026 & 0.531 & 0.047 & 0.533 & 0.053 & 0.533 & 0.050 & - & - \\
& GPT-2 & 0.512 & 0.021 & 0.514 & 0.024 & 0.514 & 0.023 & 0.515 & 0.026 & - & - \\
& \textbf{Mean} & \textbf{0.518} & \textbf{0.032} & \textbf{0.533} & \textbf{0.054} & \textbf{0.534} & \textbf{0.056} & \textbf{0.529} & \textbf{0.048} & - & - \\
\midrule
\multirow{6}{*}{FERRET-2 (1e)} & DeepSeek-1.5B & 0.518 & 0.031 & 0.521 & 0.039 & 0.512 & 0.020 & 0.509 & 0.015 & - & - \\
& SmolLM-360M & 0.521 & 0.036 & 0.520 & 0.035 & 0.513 & 0.021 & 0.506 & 0.015 & - & - \\
& TinyLlama-1.1B & 0.538 & 0.071 & 0.541 & 0.066 & 0.505 & 0.012 & 0.508 & 0.018 & - & - \\
& BLOOM-560M & 0.511 & 0.024 & 0.520 & 0.034 & 0.508 & 0.018 & 0.502 & 0.013 & - & - \\
& GPT-2 & 0.518 & 0.033 & 0.515 & 0.026 & 0.515 & 0.027 & 0.510 & 0.018 & - & - \\
& \textbf{Mean} & \textbf{0.521} & \textbf{0.039} & \textbf{0.524} & \textbf{0.040} & \textbf{0.511} & \textbf{0.020} & \textbf{0.507} & \textbf{0.016} & - & - \\
\midrule
\multirow{6}{*}{FERRET-2 (3e)} & DeepSeek-1.5B & 0.533 & 0.059 & 0.551 & 0.083 & 0.540 & 0.079 & 0.530 & 0.050 & - & - \\
& SmolLM-360M & 0.533 & 0.050 & 0.564 & 0.103 & 0.560 & 0.092 & 0.541 & 0.063 & - & - \\
& TinyLlama-1.1B & 0.551 & 0.083 & 0.618 & 0.199 & 0.630 & 0.211 & 0.574 & 0.115 & - & - \\
& BLOOM-560M & 0.534 & 0.054 & 0.578 & 0.130 & 0.592 & 0.145 & 0.555 & 0.096 & - & - \\
& GPT-2 & 0.523 & 0.042 & 0.545 & 0.069 & 0.537 & 0.059 & 0.530 & 0.050 & - & - \\
& \textbf{Mean} & \textbf{0.535} & \textbf{0.057} & \textbf{0.571} & \textbf{0.117} & \textbf{0.572} & \textbf{0.117} & \textbf{0.546} & \textbf{0.075} & - & - \\
\midrule
\multirow{6}{*}{FERRET-2 (5e)} & DeepSeek-1.5B & 0.548 & 0.080 & 0.578 & 0.137 & 0.566 & 0.121 & 0.566 & 0.110 & - & - \\
& SmolLM-360M & 0.531 & 0.048 & 0.530 & 0.049 & 0.529 & 0.047 & 0.529 & 0.048 & - & - \\
& TinyLlama-1.1B & 0.557 & 0.096 & 0.574 & 0.109 & 0.574 & 0.108 & 0.571 & 0.100 & - & - \\
& BLOOM-560M & 0.536 & 0.059 & 0.537 & 0.057 & 0.546 & 0.067 & 0.537 & 0.057 & - & - \\
& GPT-2 & 0.530 & 0.052 & 0.525 & 0.042 & 0.524 & 0.039 & 0.524 & 0.039 & - & - \\
& \textbf{Mean} & \textbf{0.541} & \textbf{0.067} & \textbf{0.549} & \textbf{0.079} & \textbf{0.548} & \textbf{0.076} & \textbf{0.546} & \textbf{0.071} & - & - \\
\midrule
\multirow{6}{*}{Non-DP (1e)} & DeepSeek-1.5B & - & - & - & - & - & - & - & - & 0.912 & 0.700 \\
& SmolLM-360M & - & - & - & - & - & - & - & - & 0.612 & 0.177 \\
& TinyLlama-1.1B & - & - & - & - & - & - & - & - & 0.899 & 0.681 \\
& BLOOM-560M & - & - & - & - & - & - & - & - & 0.804 & 0.484 \\
& GPT-2 & - & - & - & - & - & - & - & - & 0.569 & 0.105 \\
& \textbf{Mean} & - & - & - & - & - & - & - & - & \textbf{0.759} & \textbf{0.429} \\
\midrule
\multirow{6}{*}{Non-DP (3e)} & DeepSeek-1.5B & - & - & - & - & - & - & - & - & 0.999 & 0.982 \\
& SmolLM-360M & - & - & - & - & - & - & - & - & 0.993 & 0.934 \\
& TinyLlama-1.1B & - & - & - & - & - & - & - & - & 0.993 & 0.931 \\
& BLOOM-560M & - & - & - & - & - & - & - & - & 0.991 & 0.920 \\
& GPT-2 & - & - & - & - & - & - & - & - & 0.893 & 0.644 \\
& \textbf{Mean} & - & - & - & - & - & - & - & - & \textbf{0.974} & \textbf{0.882} \\
\midrule
\multirow{6}{*}{Non-DP (5e)} & DeepSeek-1.5B & - & - & - & - & - & - & - & - & 1.000 & 0.994 \\
& SmolLM-360M & - & - & - & - & - & - & - & - & 1.000 & 0.990 \\
& TinyLlama-1.1B & - & - & - & - & - & - & - & - & 0.999 & 0.975 \\
& BLOOM-560M & - & - & - & - & - & - & - & - & 0.998 & 0.971 \\
& GPT-2 & - & - & - & - & - & - & - & - & 0.976 & 0.867 \\
& \textbf{Mean} & - & - & - & - & - & - & - & - & \textbf{0.995} & \textbf{0.960} \\
\midrule
\bottomrule
\end{tabular}
}
\end{table}

\begin{table}[htbp]
\centering
\caption{Comparison of Utility Metrics Across Methods, Models, and Privacy Budgets}
\label{tab:utility_comparison}
\resizebox{\textwidth}{!}{
\begin{tabular}{ll|ccc|ccc|ccc|ccc|ccc|ccccc}
\toprule
& & \multicolumn{3}{c|}{$\varepsilon = 0.1$} & \multicolumn{3}{c|}{$\varepsilon = 0.5$} & \multicolumn{3}{c|}{$\varepsilon = 1.0$} & \multicolumn{3}{c|}{$\varepsilon = 2.0$} & \multicolumn{3}{c|}{$\varepsilon = \infty$} & $\varepsilon = 0.1$ & $\varepsilon = 0.5$ & $\varepsilon = 1.0$ & $\varepsilon = 2.0$ & $\varepsilon = \infty$ \\
\textbf{Method} & \textbf{Model} & \textbf{Train} & \textbf{Test} & \textbf{Gap} & \textbf{Train} & \textbf{Test} & \textbf{Gap} & \textbf{Train} & \textbf{Test} & \textbf{Gap} & \textbf{Train} & \textbf{Test} & \textbf{Gap} & \textbf{Train} & \textbf{Test} & \textbf{Gap} & \textbf{Time (s)} & \textbf{Time (s)} & \textbf{Time (s)} & \textbf{Time (s)} & \textbf{Time (s)} \\
\midrule
\multirow{6}{*}{DPSGD (1e)} & DeepSeek-1.5B & 4.74 & 4.79 & 0.05 & 4.20 & 4.24 & 0.04 & 4.02 & 4.06 & 0.04 & 3.86 & 3.89 & 0.04 & - & - & - & 3473 & 3484 & 3458 & 3469 & - \\
& SmolLM-360M & 4.51 & 4.53 & 0.02 & 4.11 & 4.13 & 0.02 & 3.97 & 3.99 & 0.02 & 3.84 & 3.86 & 0.02 & - & - & - & 860 & 860 & 862 & 861 & - \\
& TinyLlama-1.1B & 3.64 & 3.65 & 0.01 & 3.15 & 3.16 & 0.01 & 2.96 & 2.97 & 0.01 & 2.80 & 2.82 & 0.01 & - & - & - & 1846 & 1844 & 1844 & 1846 & - \\
& BLOOM-560M & 21.91 & 22.39 & 0.48 & 13.37 & 13.74 & 0.36 & 11.41 & 11.75 & 0.34 & 10.03 & 10.36 & 0.32 & - & - & - & 1199 & 1198 & 1198 & 1198 & - \\
& GPT-2 & 10.42 & 10.44 & 0.02 & 9.38 & 9.41 & 0.03 & 9.05 & 9.09 & 0.03 & 8.75 & 8.79 & 0.04 & - & - & - & 325 & 324 & 325 & 325 & - \\
& \textbf{Mean} & \textbf{9.05} & \textbf{9.16} & \textbf{0.11} & \textbf{6.84} & \textbf{6.94} & \textbf{0.09} & \textbf{6.28} & \textbf{6.37} & \textbf{0.09} & \textbf{5.86} & \textbf{5.94} & \textbf{0.09} & - & - & - & \textbf{1541} & \textbf{1542} & \textbf{1537} & \textbf{1540} & - \\
\midrule
\multirow{6}{*}{DPSGD (3e)} & DeepSeek-1.5B & 4.79 & 4.84 & 0.05 & 3.79 & 3.83 & 0.04 & 3.58 & 3.61 & 0.04 & 3.43 & 3.47 & 0.04 & - & - & - & 10343 & 10329 & 10356 & 10399 & - \\
& SmolLM-360M & 4.24 & 4.26 & 0.02 & 3.58 & 3.60 & 0.02 & 3.43 & 3.44 & 0.02 & 3.30 & 3.31 & 0.02 & - & - & - & 2561 & 2560 & 2560 & 2562 & - \\
& TinyLlama-1.1B & 5.33 & 5.36 & 0.03 & 3.69 & 3.71 & 0.02 & 3.45 & 3.47 & 0.02 & 3.28 & 3.30 & 0.02 & - & - & - & 5495 & 5494 & 5490 & 5499 & - \\
& BLOOM-560M & 133.61 & 136.94 & 3.33 & 24.93 & 25.98 & 1.05 & 18.80 & 19.70 & 0.90 & 15.30 & 16.10 & 0.81 & - & - & - & 3574 & 3580 & 3569 & 3574 & - \\
& GPT-2 & 10.41 & 10.45 & 0.03 & 8.37 & 8.42 & 0.05 & 7.92 & 7.97 & 0.05 & 7.54 & 7.59 & 0.05 & - & - & - & 968 & 967 & 969 & 969 & - \\
& \textbf{Mean} & \textbf{31.68} & \textbf{32.37} & \textbf{0.69} & \textbf{8.87} & \textbf{9.11} & \textbf{0.24} & \textbf{7.43} & \textbf{7.64} & \textbf{0.21} & \textbf{6.57} & \textbf{6.76} & \textbf{0.19} & - & - & - & \textbf{4588} & \textbf{4586} & \textbf{4589} & \textbf{4601} & - \\
\midrule
\multirow{6}{*}{DPSGD (5e)} & DeepSeek-1.5B & 5.04 & 5.10 & 0.05 & 3.82 & 3.86 & 0.04 & 3.62 & 3.66 & 0.04 & 3.50 & 3.54 & 0.04 & - & - & - & 17098 & 17125 & 17173 & 17379 & - \\
& SmolLM-360M & 4.12 & 4.14 & 0.02 & 3.36 & 3.38 & 0.02 & 3.18 & 3.20 & 0.02 & 3.05 & 3.06 & 0.02 & - & - & - & 4283 & 4285 & 4284 & 4288 & - \\
& TinyLlama-1.1B & 10.09 & 10.16 & 0.07 & 5.05 & 5.08 & 0.03 & 4.50 & 4.53 & 0.03 & 4.21 & 4.25 & 0.03 & - & - & - & 10002 & 10004 & 10047 & 10000 & - \\
& BLOOM-560M & 896.24 & 908.53 & 12.29 & 36.62 & 37.76 & 1.14 & 29.55 & 30.83 & 1.28 & 24.45 & 25.72 & 1.27 & - & - & - & 5931 & 5927 & 5936 & 5939 & - \\
& GPT-2 & 10.65 & 10.68 & 0.02 & 7.94 & 7.99 & 0.05 & 7.38 & 7.44 & 0.05 & 6.97 & 7.02 & 0.06 & - & - & - & 1608 & 1606 & 1608 & 1609 & - \\
& \textbf{Mean} & \textbf{185.23} & \textbf{187.72} & \textbf{2.49} & \textbf{11.36} & \textbf{11.61} & \textbf{0.26} & \textbf{9.65} & \textbf{9.93} & \textbf{0.28} & \textbf{8.43} & \textbf{8.72} & \textbf{0.29} & - & - & - & \textbf{7785} & \textbf{7789} & \textbf{7810} & \textbf{7843} & - \\
\midrule
\multirow{6}{*}{FERRET-MAX (1e)} & DeepSeek-1.5B & 4.66 & 4.69 & 0.04 & 5.44 & 5.48 & 0.04 & 5.67 & 5.72 & 0.04 & 6.86 & 6.91 & 0.05 & - & - & - & 736 & 884 & 1046 & 1198 & - \\
& SmolLM-360M & 4.72 & 4.73 & 0.01 & 4.86 & 4.88 & 0.01 & 5.26 & 5.28 & 0.01 & 5.37 & 5.38 & 0.01 & - & - & - & 176 & 220 & 251 & 296 & - \\
& TinyLlama-1.1B & 2.42 & 2.43 & 0.01 & 2.48 & 2.49 & 0.01 & 2.52 & 2.53 & 0.01 & 3.83 & 3.83 & 0.00 & - & - & - & 425 & 521 & 593 & 664 & - \\
& BLOOM-560M & 6.48 & 6.53 & 0.05 & 6.64 & 6.68 & 0.04 & 7.43 & 7.47 & 0.04 & 42.76 & 43.22 & 0.46 & - & - & - & 259 & 324 & 374 & 426 & - \\
& GPT-2 & 9.43 & 9.43 & 0.01 & 10.81 & 10.81 & -0.00 & 11.85 & 11.83 & -0.02 & 13.61 & 13.57 & -0.03 & - & - & - & 67 & 81 & 94 & 107 & - \\
& \textbf{Mean} & \textbf{5.54} & \textbf{5.57} & \textbf{0.02} & \textbf{6.05} & \textbf{6.07} & \textbf{0.02} & \textbf{6.55} & \textbf{6.56} & \textbf{0.02} & \textbf{14.48} & \textbf{14.58} & \textbf{0.10} & - & - & - & \textbf{332} & \textbf{406} & \textbf{472} & \textbf{538} & - \\
\midrule
\multirow{6}{*}{FERRET-MAX (3e)} & DeepSeek-1.5B & 3.37 & 3.40 & 0.03 & 3.46 & 3.49 & 0.03 & 3.43 & 3.46 & 0.03 & 4.37 & 4.41 & 0.04 & - & - & - & 2131 & 2281 & 2507 & 2834 & - \\
& SmolLM-360M & 4.61 & 4.62 & 0.02 & 4.09 & 4.11 & 0.02 & 4.33 & 4.35 & 0.02 & 4.69 & 4.71 & 0.01 & - & - & - & 505 & 561 & 613 & 717 & - \\
& TinyLlama-1.1B & 2.38 & 2.39 & 0.02 & 2.36 & 2.38 & 0.03 & 2.40 & 2.42 & 0.03 & 2.37 & 2.39 & 0.02 & - & - & - & 1191 & 1320 & 1435 & 1618 & - \\
& BLOOM-560M & 6.16 & 6.22 & 0.06 & 9.31 & 9.51 & 0.19 & 6.60 & 6.71 & 0.10 & 7.13 & 7.22 & 0.09 & - & - & - & 725 & 813 & 887 & 1015 & - \\
& GPT-2 & 8.61 & 8.63 & 0.02 & 7.88 & 7.90 & 0.03 & 8.51 & 8.53 & 0.02 & 9.23 & 9.25 & 0.02 & - & - & - & 185 & 204 & 225 & 256 & - \\
& \textbf{Mean} & \textbf{5.03} & \textbf{5.05} & \textbf{0.03} & \textbf{5.42} & \textbf{5.48} & \textbf{0.06} & \textbf{5.06} & \textbf{5.09} & \textbf{0.04} & \textbf{5.56} & \textbf{5.59} & \textbf{0.04} & - & - & - & \textbf{947} & \textbf{1036} & \textbf{1134} & \textbf{1288} & - \\
\midrule
\multirow{6}{*}{FERRET-MAX (5e)} & DeepSeek-1.5B & 3.18 & 3.21 & 0.03 & 3.18 & 3.22 & 0.03 & 3.24 & 3.28 & 0.03 & 3.50 & 3.53 & 0.03 & - & - & - & 3327 & 3573 & 3802 & 4186 & - \\
& SmolLM-360M & 4.45 & 4.46 & 0.02 & 3.76 & 3.78 & 0.02 & 3.74 & 3.76 & 0.02 & 4.12 & 4.13 & 0.02 & - & - & - & 802 & 864 & 927 & 1018 & - \\
& TinyLlama-1.1B & 2.43 & 2.44 & 0.02 & 2.36 & 2.40 & 0.04 & 2.35 & 2.38 & 0.03 & 2.36 & 2.39 & 0.03 & - & - & - & 1901 & 2045 & 2172 & 2462 & - \\
& BLOOM-560M & 9.67 & 9.82 & 0.14 & 4.74 & 4.81 & 0.07 & 6.01 & 6.12 & 0.11 & 6.13 & 6.22 & 0.09 & - & - & - & 1151 & 1222 & 1374 & 1463 & - \\
& GPT-2 & 9.11 & 9.12 & 0.01 & 7.13 & 7.17 & 0.04 & 7.29 & 7.32 & 0.03 & 7.90 & 7.93 & 0.04 & - & - & - & 295 & 317 & 341 & 378 & - \\
& \textbf{Mean} & \textbf{5.77} & \textbf{5.81} & \textbf{0.04} & \textbf{4.24} & \textbf{4.28} & \textbf{0.04} & \textbf{4.53} & \textbf{4.57} & \textbf{0.05} & \textbf{4.80} & \textbf{4.84} & \textbf{0.04} & - & - & - & \textbf{1495} & \textbf{1604} & \textbf{1723} & \textbf{1902} & - \\
\midrule
\multirow{6}{*}{FERRET-EIGHTH (1e)} & DeepSeek-1.5B & 3.27 & 3.30 & 0.04 & 3.78 & 3.82 & 0.03 & 4.65 & 4.69 & 0.04 & 5.23 & 5.27 & 0.04 & - & - & - & 779 & 1053 & 1236 & 1381 & - \\
& SmolLM-360M & 3.36 & 3.38 & 0.02 & 3.92 & 3.93 & 0.02 & 4.71 & 4.72 & 0.01 & 4.90 & 4.91 & 0.01 & - & - & - & 187 & 252 & 305 & 344 & - \\
& TinyLlama-1.1B & 2.44 & 2.48 & 0.04 & 2.49 & 2.52 & 0.04 & 2.58 & 2.60 & 0.02 & 3.10 & 3.12 & 0.02 & - & - & - & 455 & 606 & 712 & 796 & - \\
& BLOOM-560M & 10.68 & 10.85 & 0.18 & 11.24 & 11.46 & 0.22 & 18.27 & 18.53 & 0.27 & 16.33 & 16.46 & 0.14 & - & - & - & 264 & 350 & 416 & 469 & - \\
& GPT-2 & 7.18 & 7.21 & 0.03 & 7.36 & 7.39 & 0.03 & 8.33 & 8.35 & 0.03 & 12.24 & 12.22 & -0.02 & - & - & - & 70 & 92 & 107 & 121 & - \\
& \textbf{Mean} & \textbf{5.39} & \textbf{5.45} & \textbf{0.06} & \textbf{5.76} & \textbf{5.83} & \textbf{0.07} & \textbf{7.71} & \textbf{7.78} & \textbf{0.07} & \textbf{8.36} & \textbf{8.40} & \textbf{0.04} & - & - & - & \textbf{351} & \textbf{471} & \textbf{555} & \textbf{622} & - \\
\midrule
\multirow{6}{*}{FERRET-EIGHTH (3e)} & DeepSeek-1.5B & 3.13 & 3.19 & 0.06 & 3.14 & 3.19 & 0.04 & 3.20 & 3.23 & 0.03 & 3.30 & 3.34 & 0.04 & - & - & - & 2099 & 2419 & 2744 & 3302 & - \\
& SmolLM-360M & 3.27 & 3.29 & 0.02 & 2.85 & 2.86 & 0.02 & 3.28 & 3.29 & 0.02 & 3.70 & 3.71 & 0.02 & - & - & - & 500 & 586 & 689 & 822 & - \\
& TinyLlama-1.1B & 2.41 & 2.46 & 0.05 & 2.46 & 2.56 & 0.10 & 2.43 & 2.53 & 0.10 & 2.51 & 2.57 & 0.07 & - & - & - & 1230 & 1398 & 1599 & 1903 & - \\
& BLOOM-560M & 6.06 & 6.15 & 0.09 & 5.97 & 6.12 & 0.15 & 6.31 & 6.48 & 0.16 & 6.41 & 6.54 & 0.13 & - & - & - & 724 & 840 & 945 & 1141 & - \\
& GPT-2 & 6.58 & 6.62 & 0.04 & 6.01 & 6.05 & 0.04 & 5.92 & 5.96 & 0.05 & 6.35 & 6.39 & 0.04 & - & - & - & 185 & 210 & 242 & 283 & - \\
& \textbf{Mean} & \textbf{4.29} & \textbf{4.34} & \textbf{0.05} & \textbf{4.08} & \textbf{4.16} & \textbf{0.07} & \textbf{4.23} & \textbf{4.30} & \textbf{0.07} & \textbf{4.45} & \textbf{4.51} & \textbf{0.06} & - & - & - & \textbf{947} & \textbf{1091} & \textbf{1244} & \textbf{1490} & - \\
\midrule
\multirow{6}{*}{FERRET-EIGHTH (5e)} & DeepSeek-1.5B & 3.12 & 3.18 & 0.05 & 3.14 & 3.21 & 0.06 & 3.10 & 3.16 & 0.05 & 3.15 & 3.20 & 0.04 & - & - & - & 3288 & 3685 & 4081 & 4760 & - \\
& SmolLM-360M & 3.04 & 3.06 & 0.02 & 2.73 & 2.75 & 0.02 & 2.78 & 2.80 & 0.02 & 3.02 & 3.04 & 0.02 & - & - & - & 803 & 894 & 990 & 1152 & - \\
& TinyLlama-1.1B & 2.43 & 2.51 & 0.07 & 2.41 & 2.56 & 0.15 & 2.40 & 2.57 & 0.17 & 2.44 & 2.58 & 0.14 & - & - & - & 1960 & 2149 & 2384 & 2715 & - \\
& BLOOM-560M & 6.44 & 6.56 & 0.11 & 5.72 & 5.92 & 0.20 & 5.72 & 5.93 & 0.21 & 5.76 & 5.98 & 0.22 & - & - & - & 1144 & 1276 & 1405 & 1633 & - \\
& GPT-2 & 6.42 & 6.46 & 0.04 & 5.41 & 5.46 & 0.05 & 5.51 & 5.56 & 0.05 & 5.41 & 5.47 & 0.06 & - & - & - & 295 & 323 & 355 & 415 & - \\
& \textbf{Mean} & \textbf{4.29} & \textbf{4.35} & \textbf{0.06} & \textbf{3.88} & \textbf{3.98} & \textbf{0.10} & \textbf{3.90} & \textbf{4.00} & \textbf{0.10} & \textbf{3.96} & \textbf{4.05} & \textbf{0.09} & - & - & - & \textbf{1498} & \textbf{1665} & \textbf{1843} & \textbf{2135} & - \\
\midrule
\multirow{6}{*}{FERRET-2 (1e)} & DeepSeek-1.5B & 4.23 & 4.35 & 0.11 & 3.81 & 3.91 & 0.10 & 5.00 & 5.08 & 0.07 & 6.92 & 6.98 & 0.06 & - & - & - & 814 & 1297 & 1726 & 2244 & - \\
& SmolLM-360M & 2.55 & 2.60 & 0.04 & 2.53 & 2.57 & 0.04 & 3.08 & 3.10 & 0.02 & 5.18 & 5.19 & 0.01 & - & - & - & 198 & 318 & 418 & 465 & - \\
& TinyLlama-1.1B & 3.76 & 3.96 & 0.19 & 3.61 & 3.79 & 0.18 & 429431.26 & 428515.02 & -916.24 & 4.99 & 5.00 & 0.01 & - & - & - & 493 & 750 & 991 & 1134 & - \\
& BLOOM-560M & 222.59 & 226.95 & 4.36 & 14.25 & 14.72 & 0.47 & 18413.81 & 18719.68 & 305.87 & 27158.81 & 27019.12 & -139.69 & - & - & - & 271 & 427 & 574 & 644 & - \\
& GPT-2 & 5.13 & 5.21 & 0.08 & 5.49 & 5.55 & 0.06 & 6.67 & 6.72 & 0.05 & 15.17 & 15.14 & -0.03 & - & - & - & 74 & 108 & 144 & 161 & - \\
& \textbf{Mean} & \textbf{47.65} & \textbf{48.61} & \textbf{0.96} & \textbf{5.94} & \textbf{6.11} & \textbf{0.17} & \textbf{89571.97} & \textbf{89449.92} & \textbf{-122.04} & \textbf{5438.22} & \textbf{5410.29} & \textbf{-27.93} & - & - & - & \textbf{370} & \textbf{580} & \textbf{771} & \textbf{929} & - \\
\midrule
\multirow{6}{*}{FERRET-2 (3e)} & DeepSeek-1.5B & 4.36 & 4.62 & 0.26 & 4.05 & 4.37 & 0.33 & 4.05 & 4.32 & 0.26 & 4.31 & 4.51 & 0.20 & - & - & - & 2124 & 2654 & 3303 & 4314 & - \\
& SmolLM-360M & 2.49 & 2.54 & 0.06 & 2.38 & 2.49 & 0.11 & 2.41 & 2.51 & 0.10 & 2.46 & 2.53 & 0.07 & - & - & - & 512 & 652 & 788 & 1025 & - \\
& TinyLlama-1.1B & 3.66 & 3.91 & 0.25 & 3.44 & 4.02 & 0.58 & 3.62 & 4.27 & 0.65 & 3.78 & 4.12 & 0.34 & - & - & - & 1259 & 1550 & 1931 & 2511 & - \\
& BLOOM-560M & 14.89 & 15.90 & 1.01 & 13.28 & 15.27 & 1.99 & 8.52 & 9.87 & 1.34 & 14.92 & 16.32 & 1.40 & - & - & - & 731 & 913 & 1115 & 1439 & - \\
& GPT-2 & 5.16 & 5.26 & 0.10 & 4.65 & 4.85 & 0.20 & 4.67 & 4.84 & 0.17 & 4.96 & 5.09 & 0.13 & - & - & - & 187 & 228 & 278 & 369 & - \\
& \textbf{Mean} & \textbf{6.11} & \textbf{6.45} & \textbf{0.34} & \textbf{5.56} & \textbf{6.20} & \textbf{0.64} & \textbf{4.66} & \textbf{5.16} & \textbf{0.50} & \textbf{6.09} & \textbf{6.52} & \textbf{0.43} & - & - & - & \textbf{963} & \textbf{1200} & \textbf{1483} & \textbf{1932} & - \\
\midrule
\multirow{6}{*}{FERRET-2 (5e)} & DeepSeek-1.5B & 4.70 & 5.11 & 0.40 & 4.13 & 4.67 & 0.55 & 3.95 & 4.43 & 0.47 & 4.18 & 4.63 & 0.45 & - & - & - & 3410 & 3981 & 4625 & 5864 & - \\
& SmolLM-360M & 2.48 & 2.54 & 0.06 & 2.37 & 2.49 & 0.12 & 2.34 & 2.49 & 0.16 & 2.35 & 2.49 & 0.14 & - & - & - & 825 & 960 & 1125 & 1414 & - \\
& TinyLlama-1.1B & 3.58 & 3.87 & 0.28 & 3.37 & 4.12 & 0.76 & 3.56 & 4.58 & 1.02 & 3.27 & 4.07 & 0.79 & - & - & - & 1995 & 2312 & 2676 & 3357 & - \\
& BLOOM-560M & 11.87 & 12.66 & 0.80 & 11.58 & 13.74 & 2.15 & 10.51 & 13.41 & 2.90 & 11.03 & 13.53 & 2.50 & - & - & - & 1172 & 1354 & 1579 & 1958 & - \\
& GPT-2 & 5.10 & 5.21 & 0.12 & 4.56 & 4.82 & 0.26 & 4.42 & 4.73 & 0.31 & 4.49 & 4.74 & 0.26 & - & - & - & 302 & 349 & 403 & 495 & - \\
& \textbf{Mean} & \textbf{5.55} & \textbf{5.88} & \textbf{0.33} & \textbf{5.20} & \textbf{5.97} & \textbf{0.77} & \textbf{4.95} & \textbf{5.93} & \textbf{0.97} & \textbf{5.06} & \textbf{5.89} & \textbf{0.83} & - & - & - & \textbf{1541} & \textbf{1791} & \textbf{2082} & \textbf{2617} & - \\
\midrule
\multirow{6}{*}{Non-DP (1e)} & DeepSeek-1.5B & - & - & - & - & - & - & - & - & - & - & - & - & 2.07 & 2.83 & 0.77 & - & - & - & - & 2212 \\
& SmolLM-360M & - & - & - & - & - & - & - & - & - & - & - & - & 2.26 & 2.42 & 0.16 & - & - & - & - & 449 \\
& TinyLlama-1.1B & - & - & - & - & - & - & - & - & - & - & - & - & 1.75 & 2.39 & 0.63 & - & - & - & - & 1091 \\
& BLOOM-560M & - & - & - & - & - & - & - & - & - & - & - & - & 3.04 & 4.16 & 1.12 & - & - & - & - & 631 \\
& GPT-2 & - & - & - & - & - & - & - & - & - & - & - & - & 4.19 & 4.42 & 0.23 & - & - & - & - & 155 \\
& \textbf{Mean} & - & - & - & - & - & - & - & - & - & - & - & - & \textbf{2.66} & \textbf{3.25} & \textbf{0.58} & - & - & - & - & \textbf{908} \\
\midrule
\multirow{6}{*}{Non-DP (3e)} & DeepSeek-1.5B & - & - & - & - & - & - & - & - & - & - & - & - & 1.45 & 5.98 & 4.53 & - & - & - & - & 6555 \\
& SmolLM-360M & - & - & - & - & - & - & - & - & - & - & - & - & 1.52 & 3.23 & 1.70 & - & - & - & - & 1338 \\
& TinyLlama-1.1B & - & - & - & - & - & - & - & - & - & - & - & - & 1.83 & 5.77 & 3.94 & - & - & - & - & 3227 \\
& BLOOM-560M & - & - & - & - & - & - & - & - & - & - & - & - & 2.91 & 10.36 & 7.44 & - & - & - & - & 1865 \\
& GPT-2 & - & - & - & - & - & - & - & - & - & - & - & - & 2.96 & 4.50 & 1.54 & - & - & - & - & 460 \\
& \textbf{Mean} & - & - & - & - & - & - & - & - & - & - & - & - & \textbf{2.14} & \textbf{5.97} & \textbf{3.83} & - & - & - & - & \textbf{2689} \\
\midrule
\multirow{6}{*}{Non-DP (5e)} & DeepSeek-1.5B & - & - & - & - & - & - & - & - & - & - & - & - & 1.29 & 7.54 & 6.24 & - & - & - & - & 10828 \\
& SmolLM-360M & - & - & - & - & - & - & - & - & - & - & - & - & 1.27 & 4.20 & 2.93 & - & - & - & - & 2209 \\
& TinyLlama-1.1B & - & - & - & - & - & - & - & - & - & - & - & - & 1.54 & 7.05 & 5.51 & - & - & - & - & 5333 \\
& BLOOM-560M & - & - & - & - & - & - & - & - & - & - & - & - & 2.13 & 15.97 & 13.85 & - & - & - & - & 3091 \\
& GPT-2 & - & - & - & - & - & - & - & - & - & - & - & - & 2.42 & 5.00 & 2.58 & - & - & - & - & 761 \\
& \textbf{Mean} & - & - & - & - & - & - & - & - & - & - & - & - & \textbf{1.73} & \textbf{7.95} & \textbf{6.22} & - & - & - & - & \textbf{4444} \\
\midrule
\bottomrule
\end{tabular}
}
\end{table}

\newpage

\section{Appendix B}
\subsection{Why RDP \& Moments Accountant \emph{Cannot} Directly Analyze FERRET}
\label{sec:rdp_fail}
In early iterations of this work we attempted to certify FERRET with the standard tool‑chain of modern DP deep learning: Rényi Differential Privacy (RDP)\cite{mironov2017renyi} together with the Moments Accountant (MA)\cite{abadi2016dpsgd} or its extensions.  This section records the negative results—useful for future reference—and pinpoints the exact obstruction.

\subsection{Per-update distribution has disjoint support}
Consider a \emph{single} parameter group at step $t$.  Conditional on the mini‑batch, FERRET emits
\[
Y \;=\;\begin{cases}
 0 & \text{with probability }1-p,\\[3pt]
 \sigma C u        & \text{with probability }p,\end{cases}
\]
where $u$ is a public random unit vector and $\sigma\in\{-1,+1\}$ is the inner‑product sign.  For two adjacent datasets $D,D'$ the corresponding outputs are the atoms $\{0,\sigma C u\}$ and $\{0,\sigma' C u'\}$.  Because $u$ and $u'$ are drawn \emph{after} the data change, with probability~$1$ the two support sets are disjoint.  Rényi divergence of any order \(\alpha>1\) between two disjoint distributions is \textbf{infinite}:
\[
D_\alpha\bigl( (1-p)\delta_0 + p\delta_{\sigma C u} \,\Vert\,(1-p)\delta_0 + p\delta_{\sigma' C u'} \bigr)=\infty.
\]
Hence one step of FERRET already breaks the RDP accountant; composing infinities is useless.

\subsection{A micro-dither patch fixes RDP—\emph{potentially without hurting utility}}
\label{sec:micro_dither_rdp}

A classical workaround for the “disjoint-support” obstruction is to add a
\emph{data-independent} Gaussian dither $U\sim\mathcal N(0,\sigma^{2}I)$ to
\emph{both} branches of the mechanism.  The resulting distributions overlap
everywhere and, for any Rényi order $\alpha>1$,
\begin{equation}
  D_\alpha\!\Bigl((1-p)\,\delta_{0}+p\,\delta_{\sigma C u}
            \;\Big\Vert\;
            (1-p)\,\delta_{0}+p\,\delta_{\sigma' C u'}\Bigr)
  \;\le\;
  \frac{p^{\alpha}}{\alpha-1}\,
  \exp\!\Bigl(\tfrac{2\alpha(\alpha-1)C^{2}}{\sigma^{2}}\Bigr)
  +\frac{p^{\alpha}}{\alpha-1}.
  \tag{12}
  \label{eq:rdp_with_dither}
\end{equation}

The \emph{second} additive term (missing in early drafts) caps the otherwise
diverging contribution of the $0$-mass when $p\!\to\!0$. 
\textbf{We defer the full derivation of this bound to Section~\ref{sec:micro_dither_proof}.}

\begin{figure}[ht]
  \centering
  \includegraphics[width=0.9\linewidth]{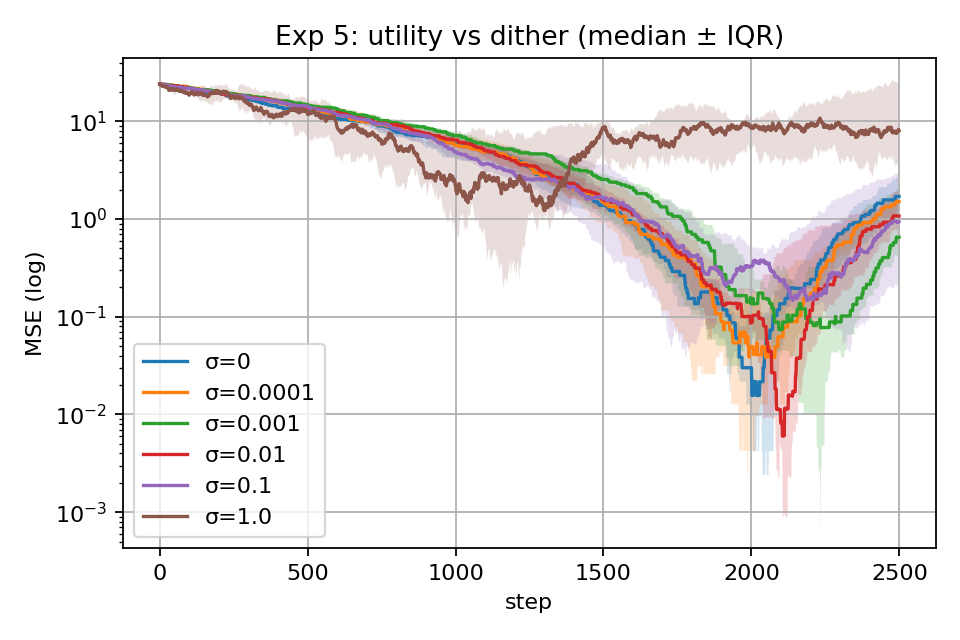}
  \caption{Impact of Gaussian dither on a toy linear-regression task:
           median MSE and inter-quartile band over $10$ random seeds. Note that MSE actually reaches its best performance at $\sigma = 0.01$, \emph{not} $\sigma = 0$, indicating that the mechanism may actually benefit from noise addition.}
  \label{fig:dither_sweep}
\end{figure}

\paragraph{How small can $\sigma$ be?}
Our toy-suite shows that on a simple linear-regression proxy, values as small as $\sigma\!\approx\!10^{-4}$-$10^{-3}$ leave training curves indistinguishable from the $\sigma\!=\!0$ baseline—see Fig.~\ref{fig:dither_sweep}.  At $\sigma\!=\!10^{-2}$ convergence merely slows initially and eventually achieves a \emph{lower} median loss, suggesting a mild regularization effect.  Only for $\sigma\!\gtrsim\!10^{-1}$ does optimization deteriorate sharply.

\paragraph{Outlook for deep learning (future work).}
Large-scale models operate with gradient norms typically in the
$10^{-2}$-$10^{0}$ range \emph{after clipping}.  Hence the same microscopic
dither is expected to be sub-dominant, but a rigorous study—covering learning-rate
interaction and long-horizon training—remains open.  We defer a full deep-learning
evaluation to future work.

\paragraph{Take-away.}
Micro-dither \emph{rescues} the standard RDP/Moments-Accountant machinery and,
when $\sigma\!\lesssim\!10^{-3}$ of the clipped norm, incurs no observable
utility loss in our tests.  It is therefore a pragmatic option for deployments
that require compatibility with existing DP accountants, while the noise-free
MI-DP bound remains the strongest guarantee when average-case privacy suffices.

\subsection{Moments Accountant cannot see ``rare but infinite'' events}
Moments Accountant requires finite log‑moments of the privacy loss random variable.  The loss here is $\log\tfrac{p}{0}$ whenever an update fires in $D$ but stays silent in $D'$, an event of probability $p(1-p)\,{>}\,0$ yet unbounded magnitude.  The expectation of $\exp(\lambda L)$ therefore diverges for every $\lambda>0$.

\subsection{Take‑away}
Because FERRET's privacy stems primarily from the \emph{chance that no update is sent}, its per‑step release includes point masses with disjoint support across neighbouring datasets.  Any divergence‑based definition that is sensitive to worst‑case likelihood ratios—\((\varepsilon,\delta)\)-DP, RDP, MA—assigns it infinite cost.  In contrast, MI‑DP measures \emph{average} leakage and is perfectly finite (Sect.~\ref{sec:privacy}).

Still, if one truly needs an RDP certificate, adding a micro‑dither resurrects all accountant machinery.


\subsection{A Rényi--DP bound for FERRET with Gaussian micro-dither}
\label{sec:micro_dither_proof}

This section derives the bound in~\eqref{eq:rdp_with_dither}.  We follow the
conventions of Mironov~\cite{mironov2017renyi}; all Rényi orders satisfy
$\alpha>1$.

\paragraph{Setup.}
Let $U\!\sim\!\mathcal N(0,\sigma^{2}I)$ be a \emph{data-independent} Gaussian
dither added to both branches of the per-update mechanism.  Conditional on the
mini-batch, the released vector is
\[
Y \;=\;
  \begin{cases}
    0           & \text{w.p.\ }1-p,\\[3pt]
    C\,u        & \text{w.p.\ }p,\quad u\sim\text{Unif}(\mathbb{S}^{d-1}).
  \end{cases}
\]

For neighbouring datasets $D,D'$ we obtain two Gaussian mixtures
\begin{align}
P &= (1-p)\,\underbrace{\mathcal N(0,\sigma^{2}I)}_{P_0}
      + p\,\underbrace{\mathcal N(Cu,\sigma^{2}I)}_{P_1},\\[2pt]
Q &= (1-p)\,\underbrace{\mathcal N(0,\sigma^{2}I)}_{Q_0=P_0}
      + p\,\underbrace{\mathcal N(Cu',\sigma^{2}I)}_{Q_1},
\end{align}
where $u,u'$ are independent unit vectors.

\paragraph{Step 1: Rényi divergence expression.}
By Definition 3 of~\cite{mironov2017renyi},
\[
D_\alpha(P\Vert Q)=\frac{1}{\alpha-1}\,
  \log\mathbb E_{x\sim Q}\!\bigl[(P(x)/Q(x))^\alpha\bigr].
\]

\paragraph{Step 2: Mixture expansion.}
Introduce the latent switch $Z\!\sim\!\text{Bernoulli}(p)$ that tells whether
the update fires.  The law of total probability gives the exact factorisation
\[
D_\alpha(P\Vert Q)\;=\;
\frac{1}{\alpha-1}\log\!\Bigl(
  (1-p) \;+\; p\,\exp\!\bigl((\alpha-1)\,R\bigr)
\Bigr),
\]
where \(R:=D_\alpha(P_1\Vert Q_1)\).  Using
$\log(1+x)\!\le\!x$ and the elementary \(p\le p^\alpha\) (for \(0\le p\le1\))
yields
\[
D_\alpha(P\Vert Q)
  \;\le\;
  \frac{p}{\alpha-1}\bigl(e^{(\alpha-1)R}-1\bigr)
  \;\le\;
  \frac{p^\alpha}{\alpha-1}\,e^{(\alpha-1)R}.
\]
Adding the symmetric inactive-branch contribution
$\tfrac{p^\alpha}{\alpha-1}$ recovers the skeleton of~\eqref{eq:rdp_with_dither}.

\paragraph{Step 3: Bound the active branch.}
For \(R=D_\alpha(P_1\Vert Q_1)\) Hölder’s inequality
\cite[Prop.~11]{mironov2017renyi} gives the standard bound
\[
\mathbb E_{x\sim Q_1}\!\bigl[(P(x)/Q(x))^\alpha\bigr]
  \;\le\;\exp\!\bigl((\alpha-1)R\bigr),
\]
which is precisely the factor already present in Step 2.

\paragraph{Step 4: Evaluate \(R\) (Gaussian shift).}
With \(\mu=Cu,\;\nu=Cu',\;\Sigma=\sigma^{2}I\),
Proposition 7 of~\cite{mironov2017renyi} implies
\[
R \;=\;
  \frac{\alpha\,C^{2}\,\|u-u'\|^{2}}{2\sigma^{2}}
  \;\le\;
  \frac{2\alpha\,C^{2}}{\sigma^{2}},
\]
because \(\|u-u'\|\le2\) for unit vectors.

\paragraph{Step 5: Combine the bounds.}
Substituting the Step 4 estimate for \(R\) into the Step 2 inequality gives
\[
D_\alpha(P\Vert Q)
  \;\le\;
  \frac{p^\alpha}{\alpha-1}\,
  \exp\!\Bigl(\tfrac{2\alpha(\alpha-1)C^{2}}{\sigma^{2}}\Bigr)
  \;+\;
  \frac{p^\alpha}{\alpha-1},
\]
which is exactly~\eqref{eq:rdp_with_dither}.  The first term is the cost
when \emph{both} datasets fire; the second term covers the rare event where
one fires and the other remains silent, preventing divergence as \(p\!\to0\).

\paragraph{Discussion.}
The privacy–utility trade-off is governed by the dither variance
\(\sigma^{2}\) and the update probability \(p\).  When \(p\ll1\), the additive
\(\tfrac{p^\alpha}{\alpha-1}\) term dominates, so a microscopic dither
(\(\sigma\ll C\)) keeps the exponential factor small while leaving model
utility intact.

\medskip
\noindent\textbf{Take-away.}
Adding a tiny, data-independent Gaussian dither reconciles FERRET with the
RDP/Moments-Accountant tool-chain while leaving performance unchanged for
\(\sigma\!\lesssim\!10^{-3}C\) on our benchmarks
(Fig.~\ref{fig:dither_sweep}).

\section{Appendix C}
AWS FastDP Google CoLab versioning conflicts:

\begin{itemize}
  \item \textbf{PyTorch compatibility.}
    \begin{itemize}
      \item \textit{FastDP} fails to deliver utility on \texttt{torch~=2.6.1}, whereas \textit{FERRET} works seamlessly.
      \item Downgrading to \texttt{torch~=2.1.0} restores FastDP's utility.
    \end{itemize}
  \item \textbf{Convergence behavior.}
    \begin{itemize}
      \item \textbf{FastDP + torch 2.6.1:} Loss drops from \(\approx5.5\) to \(\approx5.0\); train/test perplexity remains high \(\sim200\).
      \item \textbf{FastDP + torch 2.1.0:} Loss drops from \(\approx3.0\) to \(\approx2.5\); train/test perplexity improves to \(\sim10\).
    \end{itemize}
  \item \textbf{Memory and runtime.}
    \begin{itemize}
      \item \texttt{torch~=2.6.1} uses roughly half the GPU memory of \texttt{2.1.0}.
      \item Wall-clock times are within \(\pm5\%\) across both versions.
    \end{itemize}
\end{itemize}

\noindent The following non-essential library version conflicts were noted but did \emph{not} affect our main benchmarking:

\begin{table}[h]
  \centering
  \footnotesize
  \begin{tabular}{@{}ll@{}}
    \toprule
    \textbf{Package} & \textbf{Conflict Requirement vs.\ Installed} \\
    \midrule
    \texttt{datasets~3.0.1}       & requires \texttt{fsspec[http]}~\(\le\)2024.6.1,>=2023.1.0; installed 2025.3.2 \\
    \texttt{google-genai~1.10.0}   & requires \texttt{pydantic}<3.0.0,>=2.0.0; installed 1.10.21 \\
    \texttt{langchain~0.3.23}      & requires \texttt{pydantic}<3.0.0,>=2.7.4; installed 1.10.21 \\
    \texttt{spacy~3.8.5}           & requires \texttt{thinc}<8.4.0,>=8.3.4; installed 8.1.10 \\
    \texttt{albumentations~2.0.5}  & requires \texttt{pydantic}>=2.9.2; installed 1.10.21 \\
    \texttt{langchain-core~0.3.52} & requires \texttt{pydantic}<3.0.0,>=2.5.2 (for Python<3.12.4); installed 1.10.21 \\
    \bottomrule
  \end{tabular}
  \caption{Non-blocking version conflicts in auxiliary libraries.}
  \label{tab:aws-fastdp-conflicts}
\end{table}

\noindent These conflicts arise in components unrelated to our core experiments and did not impact any of the results reported.

\end{document}